# Likelihood based inference for high-dimensional extreme value distributions

Alexis BIENVENÜE and Christian Y. ROBERT

Université de Lyon, Université Lyon 1, Institut de Science Financière et d'Assurances,
50 Avenue Tony Garnier, F-69007 Lyon, France

December 30, 2014

## Abstract

Multivariate extreme value statistical analysis is concerned with observations on several variables which are thought to possess some degree of tail-dependence. In areas such as the modeling of financial and insurance risks, or as the modeling of spatial variables, extreme value models in high dimensions (up to fifty or more) with their statistical inference procedures are needed. In this paper, we consider max-stable models for which the spectral random vectors have absolutely continuous distributions. For random samples with max-stable distributions we provide quasi-explicit analytical expressions of the full likelihoods. When the full likelihood becomes numerically intractable because of a too large dimension, it is however necessary to split the components into subgroups and to consider a composite likelihood approach. For random samples in the max-domain of attraction of a max-stable distribution, two approaches that use simpler likelihoods are possible: (i) a threshold approach that is combined with a censoring scheme, (ii) a block maxima approach that exploits the information on the occurrence times of the componentwise maxima. The asymptotic properties of the estimators are given and the utility of the methods is examined via simulation. The estimators are also compared with those derived from the pairwise composite likelihood method which has been previously proposed in the spatial extreme value literature.

## 1 Introduction

Technological innovations have had deep impact on scientific research by allowing to collect massive amount of data with relatively low cost. These new data create opportunities for the development of data analysis. In particular the availability of high-dimensional data has significantly challenged the traditional multivariate statistical theory.

Problems concerning financial risks are typically of high-dimensions in character. Financial portfolios often consist of a large number of assets and complex instruments (more than one hundred) and the dependence between the underlying components of the portfolios plays a crucial role in their pricing (see e.g. [36]). Understanding the links between extreme events is also of crucial importance for an efficient quantitative risk management.

Problems involving spatial dependence are also of high-dimensions in character. Extreme environmental phenomena such as major precipitation events or extreme wind speeds manifestly exhibit spatial dependence and are of much interest for engineers because knowledge of the spatial dependence is essential for regional risk assessment.

We focus in this paper on multivariate extreme value distributions in high dimensions. Models and inference methods for these distributions is an exciting research field that is in its beginning of development. One way to characterize the dependence among multivariate extremes is to consider the tail-dependence function or the tail-dependence density derived from known copulas of multivariate distributions ([32]). Some models based on Archimedean copulas or elliptical copulas have been proposed (see e.g. [28], [30]) but no general inference method has been associated to these models. The literature on modeling high-dimensional extreme events mainly concerns spatial max-stable processes. These processes arise as limits in distribution of componentwise maxima of independent spatial processes, under suitable centering and normalization. Max-stable processes are very useful to study extreme value phenomena for which spatial patterns can be discerned.

Likelihood inference for Archimedean copulas in high dimensions has been recently investigated in [23] and gives an opportunity for the only Archimedean extreme value copula, the Logistic (also known as Gumbel–Hougaard) copula. But this approach is clearly insufficient because it is limited to exchangeable random variables whose dependence is characterized by this Archimedean copula.

The composite likelihood methods that are based on combinations of valid likelihood objects related to small subsets of data are rather naturally considered as an appealing way to deal with inference in high dimensions. The merit of composite likelihood is to reduce the computational complexity. These methods go back to [3] and have been extensively studied in particular for the case where the full likelihood is unavailable and is replaced by quantities that combine bivariate or trivariate joint likelihoods, see e.g. [33], [7], [49]. Recent works concerning the application for (i) random samples from max-stable processes and for (ii) random samples in the max-domain of attraction of a max-stable distribution can be found in [39], [16], [24] and [45] for (i), and in [1], [27] and [31] for (ii).

For a class of max-stable processes, commonly known as Brown–Resnick processes (see [4], [29]), the full-likelihood may be derived explicitely. Extensive simulations have been performed very recently in [5] to assess the loss of information of composite likelihood estimators with respect to a full likelihood approach (the logistic model and the Reich-Shaby model for which the analytical expressions of the full likelihoods are available have also been considered). The authors have shown that high-performance computing can be a powerful tool to perform efficient inference and leads to substantial improvement in efficiency (even if dimensions should be today less than twelve due to the increasing complexity of the full likelihood with respect to the dimension). Moreover, they have observed for these models that truncation of composite likelihoods may reduce the computational burden with a very small loss of efficiency.

If random samples in the max-domain of attraction of the Brown–Resnick processes are available, [14] and [53] offer considerably more efficient strategies. [14] establishes that the multivariate conditional distribution of extremal increments with respect to a single extreme component is multivariate Gaussian, therefore offering the possibility to perform high-dimensional inference. In a recent paper ([53]), Tawn and Wadsworth exploit the limiting multivariate Poisson process intensities for inference on vectors exceeding a high marginal threshold in at least one component, employing a censoring scheme to incorporate information below the marginal threshold. They also propose to perform likelihood inference by exploiting the methods developed in [48] where information on the occurrence times of extreme events is incorporating to simplify the likelihood.

In a recent paper [25], a comparative study of likelihood estimators have been conducted in the case of the logistic model for which likelihood functions have analytical expressions. The authors have compared estimators based on the block maximum approach (full likelihood, full likelihood with occurrence times) and estimators based on the threshold-Poisson approach (with and without censoring). They have shown that threshold based estimators outperform block maximum estimators in general and that the block maxmimum estimators may be highly biased for weak independence to near independence scenarios (see [51] for a bias reduction technique).

In this paper, we consider any max-stable distribution whose spectral random variable has an absolutely continuous distribution.

For random samples with a multivariate max-stable distribution, we provide quasi-explicit analytical expressions of the full likelihood. When this likelihood becomes numerically intractable, we gather together components of the vector within clusters with bounded sizes for which the likelihood is feasible with a moderate computational cost and we consider the pseudolikelihood equal to the product of the likelihoods of the clusters.

For random samples in the max-domain of attraction, we propose two approaches that simplify a lot the full likelihood. First we consider the multivariate density function of the vectors of exceedances (for which at least one component exceed a hight threshold) rather than the intensities of the Poisson process like in [53]. Second we use the joint density function of the componentwise maxima over blocks and of the partition variable that gives which componentwise maxima occur simultaneously. As explained in [48], the likelihood is only a part of the likelihood without information on the partition in this case.

The paper is organized as follows. Section 2 gives probability density functions of multivariate max-stable distributions and probability density functions for the asymptotic distributions of the vectors of exceedances with censored components and of the vector of componentwise maxima with information on occurrences. In Section 3 we consider parametric models for random samples with max-stable distributions and for random samples in the max-domain of attraction of a max-stable distribution. We define maximum likelihood estimators and pseudo-maximum likelihood estimators built by clustering components of the vector if necessary. We prove consistency and asymptotic normality for these estimators when the dimension of the vector is fixed and as the number of observations tends to infinity. Section 4 illustrates the performance of the methods through a number of simulation studies and gives comparisons with the composite likelihood method. The proofs have been deferred to the appendix.

Our statistical procedures have been implemented in the R package HiDimMaxStable available on CRAN.

# 2 Density functions for extreme value models with absolutely continuous spectral random vectors

## 2.1 Introduction

Let $\mathbf{Y}_i = (Y_{i1}, \ldots, Y_{im})$, for $i = 1, \ldots, n$, be independent and identically distributed random vectors in dimension $m$. The analysis of multivariate extreme value values has first been based on the weak convergence of the vector of componentwise maxima $\mathbf{M}_n = (M_{n1}, \ldots, M_{nm})$ where $M_{nj} = \max_{i=1,\ldots,n} Y_{ij}$. Under general conditions $\mathbf{M}_n$, suitably normalized, converges in distribution to a member of the multivariate extreme value distributions (see e.g. [41]). Suppose that the marginal components of the multivariate extreme value distribution $G_*$ have a unit Fréchet distribution, then there exists a random vector $\mathbf{U} = (U_1, \ldots, U_m)$ on $\mathbb{R}^m$ with $\mathbb{E}[U_j^+] = 1$ for $j = 1, \ldots, m$ where $U_j^+ = \max(0, U_j)$, such that the tail-dependence function of $G_*$, defined as

$V_*(\mathbf{z}) = -\log\left(G_*(\mathbf{z})\right)$ for $\mathbf{z} = (z_1, ..., z_m) \in \mathbb{R}_+^m$, is given by

$$V_*(\mathbf{z}) = \mathbb{E}\left[\max\left(z_1^{-1} U_1^+, \ldots, z_m^{-1} U_m^+\right)\right]. \tag{2.1}$$

Moreover it may be shown that, if $\mathbf{Z}$ has distribution $G_*$, then

$$\mathbf{Z} \stackrel{d}{=} \max_{j \geq 1} \zeta_j \mathbf{U}_j,$$

where $(\zeta_j)_{j \geq 1}$ are the points of a Poisson process on $\mathbb{R}^+$ with intensity $ds/s^2$, and $(\mathbf{U}_j)_{j \geq 1}$ is a sequence of independent and identically distributed random vectors with the same distribution as $\mathbf{U}$ (see Corollary 9.4.5 and Remark 9.4.8 in [11]). The random vector $\mathbf{U}$ is referred to as the spectral random vector of $G_*$ (or sometimes as the profile vector).

Such tail-dependence functions appear for the finite dimensional distributions of simple max-stable processes characterized by Schlather's spectral representation (see [46] or [40] where a representation of the same type for the (semi-)min-stable case is given). There are few spatial models for which $V_*(\mathbf{z})$ is analytically known for dimension $m$ larger than two. Exceptions are given by (i) the spatial max-stable processes for which the spectral random vectors are Log-normal (e.g. the Geometric Gaussian max-stable process and the Brown-Resnick max-stable process; [24] proves that $V_*(\mathbf{z})$ may be written as a linear combination of $(m-1)$-dimensional multivariate normal probabilities, see also [27]), and (ii) the spatial max-stable processes for which the spectral random vectors are equal to a power (of the positive part) of a Gaussian random vector (e.g. the Extremal t max-stable process, see [38]); [37] proves that $V_*(\mathbf{z})$ may be written as a linear combination of $(m-1)$-dimensional multivariate Student probabilities.

When the $U_j$ are independent and identically distributed random variables, it is possible to derive several well-known analytical forms for the tail-dependence functions. Even if they lead to distributions of exchangeable random variables, such tail dependence functions may offer interest for inference and practical applications. Let $\eta_j$, for $j = 1, ..., m$, be positive independent and identically distributed random variables with distribution function $H$, such that $\mathbb{E}[\eta_j] < \infty$. Let $U_j = \eta_j / \mathbb{E}[\eta_j]$. Simple calculations give

$$V_*(\mathbf{z}) = \sum_{l=1}^m z_l^{-1} \mathbb{E}\left[\frac{\eta}{\mathbb{E}[\eta]} \prod_{s \neq l} H\left(\left(\frac{z_s}{z_l}\right)\eta\right)\right]$$

where $\eta$ is a positive random variable with distribution $H$. If $H$ is a Bernoulli distribution, then $G_*$ is the Marshall-Olkin distribution [34]. If $H$ is a Weibull distribution, then $G_*$ is the Galambos distribution [15]. If $H$ is a Fréchet distribution, then $G_*$ is the logistic (or the Gumbel) distribution [22].

## 2.2 Probability density functions of multivariate max-stable distributions

Assume that $\mathbf{Z} = (Z_1, \ldots, Z_m)$ has a max-stable distribution with tail-dependence function $V_*$. If the $m$-th order partial derivatives of $V_*$ exist, then the density function of $\mathbf{Z}$ may be derived by Faà di Bruno's formula for multivariate functions. Let $\mathcal{I} = \{1, \ldots, m\}$. We denote by $\Pi$ the set of all partitions of $\mathcal{I}$. For a partition $\pi \in \Pi$, $B \in \pi$ means that $B$ is one of the blocks of the partition $\pi$. For a set $B \subset \mathcal{I}$, we let $\mathbf{z}_B = (z_j)_{j \in B}$. The density function of $\mathbf{Z}$, $h(\mathbf{z})$, is then given by

$$h(\mathbf{z}) = \exp\left(-V_*(\mathbf{z})\right) \sum_{\pi \in \Pi} (-1)^{|\pi|} \prod_{B \in \pi} \frac{\partial^{|B|}}{\partial \mathbf{z}_B} V_*(\mathbf{z})$$

where $|\pi|$ denotes the number of blocks of the partition $\pi$, $|B|$ denotes the cardinality of the set $B$ and $\partial^{|B|}/\partial \mathbf{z}_B$ denotes the partial derivative with respect to $\mathbf{z}_B$ (see also [48], [43], [5], [25]).

We now assume that $\mathbf{Z}$ has a max-stable distribution whose spectral random vector $\mathbf{U}$ has an absolutely continuous distribution. For a set $B \subset \mathcal{I}$, we let $\{\mathbf{U}_B \leq \mathbf{z}_B\} = \cap_{j \in B}\{U_j \leq z_j\}$ and define

$$\mu\left(B; \mathbf{z}\right) = \int_0^\infty \gamma^{|B|} \Pr\left(\mathbf{U}_{B^c} \leq \mathbf{z}_{B^c}\gamma \,|\, \mathbf{U}_B = \mathbf{z}_B\gamma\right) f_{\mathbf{U}_B}\left(\mathbf{z}_B\gamma\right) d\gamma \qquad (2.2)$$

where $B^c = \mathcal{I}\backslash B$ and $f_{\mathbf{U}_B}$ is the probability density function of $\mathbf{U}_B$. Note that, if the conditional distribution of $\mathbf{U}_{B^c}$ given $\mathbf{U}_B$ and the density functions $f_{\mathbf{U}_B}$ are analytically known, then $\mu\left(B; \mathbf{z}\right)$ may be easily computed because it is a one-dimensional integral. If the $U_j$ are independent and identically distributed random variables, it is possible to get analytical expressions for the $\mu\left(B; \mathbf{z}\right)$ (for example when $U_j$ has a Weibull or a Fréchet distribution (leading to the logistic model)).

**Proposition 1** *For $\mathbf{z} \in \mathbb{R}_+^m$ and $B \subset \mathcal{I}$, we have*

$$V_*(\mathbf{z}) = \sum_{l=1}^m z_l \mu\left(\{l\}\,; \mathbf{z}\right) \quad \text{and} \quad \mu\left(B; \mathbf{z}\right) = -\frac{\partial^{|B|}}{\partial \mathbf{z}_B} V_*(\mathbf{z}).$$

*It follows that*

$$h\left(\mathbf{z}\right) = \exp\left(-\sum_{l=1}^m z_l \mu\left(\{l\}\,; \mathbf{z}\right)\right)\left(\sum_{\pi \in \Pi}\prod_{B \in \pi} \mu\left(B; \mathbf{z}\right)\right). \qquad (2.3)$$

The previous proposition gives us a quasi-explicit analytical expression of the density function in the sense that it only depends on quantities $\mu\left(B; \mathbf{z}\right)$, $B \subset \mathcal{I}$. They can be efficiently computed when the spectral random vector $\mathbf{U}$ is Gaussian or Log-normal (see Appendix 6.2 for their expressions). It also the case for the following model: let $(P_i)_{i=1,\ldots,I}$ be a partition of $\mathcal{I} = \{1,\ldots,m\}$ and $\mathbf{U} = (\mathbf{U}_{P_1},\ldots,\mathbf{U}_{P_I})$ be a positive random vector in $\mathbb{R}^m$ with expectation equal to $\mathbf{e} = (1,\ldots,1)'$ such that $\mathbf{U}_{P_1},\ldots,\mathbf{U}_{P_I}$ are independent sub-vectors and

$$\Pr\left(\mathbf{U}_{P_i} \leq \mathbf{z}_{P_i}\right) = \mathrm{C}_i((F_i\left(z_j\right))_{j \in P_i})$$

where $\mathrm{C}_i$ is an Archimedean copula with a completely monotone generator $\psi_i$ and $F_i$ is the common marginal distribution function of $\mathbf{U}_{P_i}$. In that case, we have

$$\mu\left(B; \mathbf{z}\right) = \int_0^\infty \gamma^{|B|} \prod_{i=1}^I \left[\psi_i^{(|B \cap P_i|)}\left(\sum_{j \in P_i}\psi_i^{-1}\left(F_i\left(\gamma z_j\right)\right)\right)\prod_{j \in B \cap P_i}\frac{1}{\psi_i^{(1)}\left(\psi_i^{-1}\left(F_i\left(\gamma z_j\right)\right)\right)}f_i\left(\gamma z_j\right)\right] d\gamma \tag{2.4}$$

where $f_i$ is the probability density function of the $U_j$ for $j \in P_i$. We call the max-stable distribution associated with the tail-dependence function generated by $\mathbf{U}$ an homogeneous "clustered" max-stable distribution.

## 2.3 Asymptotic probability density functions for random vectors in the max-domain of attraction of a multivariate max-stable distribution

When considering random samples of max-stable vectors, the probability density function may be quite complex if the dimension of the vector becomes large, since it presents an explosion of terms. Indeed the number of partitions of $\mathcal{I}$ is given by the $m$-th Bell number and increases dramatically with $m$. For example $m = 7$ (resp. $m = 10$) would require to sum over around 1000 (resp. 116 000) terms. In practice, $m = 7$ is today an upper bound for with a personal computer.

If a random sample in the max-domain of attraction of the multivariate max-stable distribution is available, it should rather be used because it is possible to simplify a lot the probability density function. Following [53], two approaches may be followed.

First we may consider only the extreme events for which at least one component of the vector has exceeded a high threshold and censor the other components. The density function has then a very simple expression (note that the censored Poisson process associated with exceedances is alternatively considered in [53]).

Second we may focus on the componentwise maxima, but we also use the additional information on the partition variable that gives which componentwise maxima occur simultaneously. As explained in [48], the density function of the couple is then only an element of the density function of the (asymptotic) max-stable distribution.

**Vectors of exceedances with censored components**

Assume that $\mathbf{Y} = (Y_1, \ldots, Y_m)$ has margins with asymptotically unit Pareto distribution (i.e. satisfying $\lim_{t \to \infty} t \Pr(Y_j > ty) = y^{-1}$ for $y > 1$) and is in the max-domain of attraction of $G_*$. Hence we have

$$\lim_{t \to \infty} t \Pr\left(\cup_{j=1}^m \{Y_j > tz_j\}\right) = -\log G_*(\mathbf{z}) = \mathbb{E}[\max_{j=1,\ldots,m} z_j^{-1} U_j^+].$$

For a given threshold $t > 0$, we censor components that do not exceed the threshold and consider the vector $\mathbf{X}_t = t^{-1}\mathbf{Y} \vee \mathbf{e}$ with $\mathbf{e} = (1, \ldots, 1)' \in \mathbb{R}^m$. We are interested in the asymptotic conditional distribution of $\mathbf{X}_t^* \overset{d}{=} \mathbf{X}_t | \|\mathbf{X}_t\|_\infty > 1$ as $t$ tends to infinity, where $\|\mathbf{X}_t\|_\infty$ is the maximum norm of $\mathbf{X}_t$.

Let $\mathcal{P} = \mathcal{P}(\mathcal{I})$ be the power set of $\mathcal{I} = \{1, \ldots, m\}$ without $\varnothing$, and, for $B \in \mathcal{P}$, let $\mathcal{A}_B = \{\mathbf{z} \in [1, \infty)^m : z_i > 1, i \in B, z_i = 1, i \in B^c\}$. Note that $\mathbf{X}_t^*$ takes its values in $\mathcal{A} = \cup_{B \in \mathcal{P}} \mathcal{A}_B$. Define

$$V_B^*(\mathbf{z}) = \int_0^\infty P(\mathbf{U}_B > \gamma \mathbf{z}_B, \mathbf{U}_{B^c} \leq \gamma \mathbf{z}_{B^c}) \, d\gamma.$$

For $B \in \mathcal{P}$, let $p_B(\mathbf{z}) = V_B^*(\mathbf{z})/V^*(\mathbf{z})$. Since $\sum_{B \in \mathcal{P}} V_B^*(\mathbf{z}) = V^*(\mathbf{z})$, we deduce that $(p_B(\mathbf{z}))_{B \in \mathcal{P}}$ is a discrete probability distribution.

**Proposition 2** *As $t \to \infty$, $\mathbf{X}_t^*$ converges in distribution to $\mathbf{X}^*$ whose density function $f_{\mathbf{X}^*}$ is characterized in the following way: for $B \in \mathcal{P}$ and $\mathbf{x} = (\mathbf{x}_B, \mathbf{e}_{B^c}) \in \mathcal{A}_B$*

$$f_{\mathbf{X}^*(\mathcal{A}_B)}(\mathbf{x}_B) = \frac{\mu(B; \mathbf{x})}{V_B^*(\mathbf{e})}$$

*where $f_{\mathbf{X}^*(\mathcal{A}_B)}$ is the conditional density function of $\mathbf{X}^*$ given that $\mathbf{X}^* \in \mathcal{A}_B$. It follows that, for $\mathbf{x} \in \mathcal{A}$,*

$$f_{\mathbf{X}^*}(\mathbf{x}) = \sum_{B \in \mathcal{P}} p_B(\mathbf{e}) f_{\mathbf{X}^*(\mathcal{A}_B)}(\mathbf{x}_B) \mathbb{I}_{\{\mathbf{x} \in \mathcal{A}_B\}} = \frac{1}{\sum_{i=1}^m \mu(\{i\}; \mathbf{e})} \sum_{B \in \mathcal{P}} \mu(B; \mathbf{x}) \mathbb{I}_{\{\mathbf{x} \in \mathcal{A}_B\}}.$$

Note that the distribution of $\mathbf{X}^*$ is not a multivariate Generalized Pareto distribution as introduced in [44] because the components of $\mathbf{Y}$ that do not exceed the threshold have been censored.

We only focus on extreme components and do not want to suffer of misspecification below the threshold.

**Componentwise maxima with additional information about maxima occurrences**

Let $\mathbf{Y}_i = (Y_{i1}, \ldots, Y_{im})$, for $i = 1, \ldots, n$, be independent and identically distributed random vectors whose margins have (asymptotically) unit Pareto distributions and that are in the max-domain of attraction of $G_*$. The vector of componentwise maxima, $\mathbf{M}_n$, is such that

$$\Pr(\mathbf{M}_n \leq n\mathbf{z}) = F_{\mathbf{Y}}^n(n\mathbf{z}) \to \exp(-V_*(\mathbf{z})).$$

We further incorporate the information about the partition $\mathbf{R}_n$ that gives which componentwise maxima occur simultaneously. For example, for $m = 3$ and $\mathbf{M}_n = (M_{1,n}, M_{2,n}, M_{3,n})$, if $M_{2,n}$ and $M_{3,n}$ occured simultaneously, but separately from $M_{1,n}$, then $\mathbf{R}_n = \{1, \{2, 3\}\}$.

**Proposition 3** *As $n \to \infty$, $(n^{-1}\mathbf{M}_n, \mathbf{R}_n)$ converges in distribution to $(\mathbf{M}, \mathbf{R})$ whose density function $f_{(\mathbf{M}, \mathbf{R})}$ is characterized in the following way: for $\pi \in \Pi$ and $\mathbf{z} \in \mathbb{R}_+^m$*

$$f_{(\mathbf{M}, \mathbf{R}=\pi)}(\mathbf{z}) = \exp\left(-\sum_{l=1}^m z_l \mu\left(\{l\}; \mathbf{z}\right)\right) \left(\prod_{B \in \pi} \mu\left(B; \mathbf{z}\right)\right).$$

The proof is omitted since the main arguments may be found in [48].

# 3 Likelihood inference

## 3.1 Assumptions

We impose that the probability density function of $\mathbf{U}$ belongs to some parametric family $\{f_{\mathbf{U}}(\cdot; \theta), \theta \in \Theta\}$ where $\Theta$ is a compact set in $\mathbb{R}^p$, for $p \geq 1$, and let

$$\mu(\theta; B, \mathbf{z}) = \int_0^\infty \int_{-\infty}^{\mathbf{z}_{B^c}} \gamma^m f_{\mathbf{U}_{B^c}, \mathbf{U}_B}(\mathbf{u}_{B^c}\gamma, \mathbf{z}_B\gamma; \theta) \, d\mathbf{u}_{B^c} d\gamma.$$

For the existence of $\nabla_\theta \mu(\theta; B, \mathbf{z})$ and $\nabla_\theta^2 \mu(\theta; B, \mathbf{z})$, we will assume the following conditions:

- $C1$: There exist gradient functions $\nabla_\theta \log f_{\mathbf{U}}(\mathbf{z}; \theta)$, Hessian functions $\nabla_\theta^2 \log f_{\mathbf{U}}(\mathbf{z}; \theta)$ and $\nabla_\theta^3 \log f_{\mathbf{U}}(\mathbf{z}; \theta)$ for any $\mathbf{z} \in \mathbb{R}_+^m$.

- $C2$: It is possible to interchange differentiation (with respect to $\theta$) and integration (with respect to $\gamma$ and $\mathbf{z}_{B^c}$) for

$$\int_0^\infty \int_{-\infty}^{z_{B^c}} \gamma^m \nabla_\theta^j f_{\mathbf{U}_{B^c}, \mathbf{U}_B}(\mathbf{u}_{B^c}\gamma, \mathbf{z}_B\gamma; \theta) \, d\mathbf{u}_{B^c} d\gamma \quad j = 0, 1, 2.$$

If $\mathbf{Y} = (Y_1, \ldots, Y_m)$ is in the max-domain of attraction of $G_*$, the following condition could also be used:

- $C3$: There exists $\alpha > 0$ such that, uniformly for $B \in \mathcal{P}$ and $\mathbf{x} = (\mathbf{x}_B, \mathbf{e}_{B^c}) \in \mathcal{A}_B$, as $t \to \infty$,

$$\Pr\left(\mathbf{X}_{B,t}^* \in (\mathbf{e}_B, \mathbf{x}_B] | \mathbf{X}_t^* \in \mathcal{A}_B\right) - F_{\mathbf{X}^*(\mathcal{A}_B)}(\mathbf{x}_B) = O\left(t^{-\alpha}\right),$$

where $F_{\mathbf{X}^*(\mathcal{A}_B)}$ is the multivariate distribution function of $\mathbf{X}^*$ restricted to $\mathcal{A}_B$ (defined in Proposition 2).

## 3.2 Likelihood inference for random samples of high dimensional max-stable distributions

We consider an identifiable statistical model: $\mathcal{F} = \{h(\theta; \mathbf{z}_i), \theta \in \Theta, \mathbf{z}_i \in \mathbb{R}_+^m, i = 1, \ldots, n\}$ where $\mathbf{z}_i$ are observations of a sample of size $n$ distributed as $\mathbf{Z} = (Z_1, \ldots, Z_m)$ with density function $h(\theta; \cdot)$ as in Proposition 1. Let $\theta_0$ denote the true parameter. We assume that $\theta_0$ is interior to the parameter space $\Theta$.

The log-likelihood of an observation $\mathbf{z} \in \mathbb{R}_+^m$ is given by

$$\ell_1(\theta; \mathbf{z}) = -\sum_{l=1}^m z_l \mu(\theta; \{l\}, \mathbf{z}) + \log\left(\sum_{\pi \in \Pi} \prod_{B \in \pi} \mu(\theta; B, \mathbf{z})\right).$$

As explained previously, the computation of this log-likelihood requires to consider the set $\Pi$ of all possible partitions of the components of $\mathbf{z}$. Due to the explosive behavior of the number of partitions, this likelihood may become numerically intractable for a too large number of components. We therefore decide to also consider the partition-composite likelihood

$$\ell_2(\theta; \mathbf{z}) = \sum_{B \in \pi} |B| \ell_1(\theta; \mathbf{z}_B)$$

for which it is necessary to only compute the likelihoods for the blocks of a partition $\pi$. By bounding the sizes of the blocks, the computation is feasible in a moderate time. Of course, the partition should be chosen such that the components of the vector in blocks are the most dependent as possible or equivalently such that the (presumed) assumption of independence between the blocks of the partition is the most reasonable as possible. In practice, such a partition may be chosen by using a clustering algorithm like the Partitioning Around Medoids (PAM) algorithm as proposed in [2].

Moreover we will also compare the two previous log-likelihoods with the pairwise log marginal likelihood that has been first proposed by [39] for multivariate max-stable distributions

$$\ell_3(\theta; \mathbf{z}) = \sum_{i < j} \ell_1\left(\theta; \mathbf{z}_{\{i,j\}}\right)$$

and for which only $m(m-1)/2$ bivariate-likelihoods are needed.

Conditions $(C1)$ and $(C2)$ of Section 3.1 imply the existence of the score functions $\nabla_\theta \ell_j(\theta; \mathbf{z})$ with their respective Hessian functions $\nabla_\theta^2 \ell_j(\theta; \mathbf{z})$ for $j = 1, 2, 3$ (see Section 6.5 for their expressions). The maximum likelihood estimator $\hat{\theta}_n^{(1)}$ (MLE), the maximum partition-composite likelihood estimator $\hat{\theta}_n^{(2)}$ (MpcLE) and the maximum pairwise marginal likelihood estimator $\hat{\theta}_n^{(3)}$ (MpmLE) are respectively defined by the conditions

$$\sum_{i=1}^n \nabla_\theta \ell_j(\hat{\theta}_n^{(j)}; \mathbf{z}_i) = 0, \quad j = 1, 2, 3. \tag{3.5}$$

Let, for $j = 1, 2, 3$,

$$I_j(\theta) = \mathbb{E}\left[-\nabla_\theta^2 \ell_j(\theta; \mathbf{Z})\right] \quad \text{and} \quad J_j(\theta) = \mathbb{E}\left[\nabla_\theta \ell_j(\theta; \mathbf{Z}) \nabla_\theta \ell_j(\theta; \mathbf{Z})'\right].$$

We get the following proposition (given without proof, see e.g. Section 4.4.2 in [8] for $\hat{\theta}_n^{(1)}$):

**Proposition 4** *Assume that* $(C1)$ *and* $(C2)$ *hold. The likelihood estimators are consistent and asymptotically normal distributed as* $n$ *tends to infinity*

$$\sqrt{n}(\hat{\theta}_n^{(1)} - \theta_0) \xrightarrow{d} \mathcal{N}\left(0, I_1^{-1}(\theta_0)\right) \quad \text{and} \quad \sqrt{n}(\hat{\theta}_n^{(j)} - \theta_0) \xrightarrow{d} \mathcal{N}(0, I_j^{-1}(\theta_0) J_j(\theta_0) I_j^{-1}(\theta_0)), \quad j = 2, 3.$$

It is well-known that the MLE attains the Cramér–Rao lower bound $I_1^{-1}(\theta)$ and that estimation using the composite likelihood (for the MpmLE and the MpcLE) results in a loss of efficiency (see e.g. [39]). But it is very difficult to assess the level of efficiency from a theoretical point of view (see e.g. [7]) because it strongly depends on the considered model.

### 3.3 Likelihood inference for random samples in a max-domain of attraction of a multivariate max-stable distribution

Let $(\mathbf{Y}_i)_{i=1,\dots,n}$ be independent and identically vectors distributed as $\mathbf{Y} = (Y_1, \dots, Y_m)$ whose margins have (asymptotically) unit Pareto distributions and assume that $\mathbf{Y}$ is in the max-domain of attraction of $G_*$.

**Vectors of exceedances with censored components**

Let $1 \leq k \leq n$. For $i = 1, \dots, n$, we censor components that do not exceed the threshold $n/k$ and define

$$\mathbf{X}_{i,n/k} = \frac{k}{n}\mathbf{Y}_i \vee \mathbf{e}$$

where $\mathbf{e} = (1, \dots, 1)' \in \mathbb{R}^m$. We only keep random vectors $\mathbf{X}_{i,n/k}$ for which $\left\|\mathbf{X}_{i,n/k}\right\|_\infty > 1$ and let

$$\mathcal{N}_k = \left\{i = 1, \dots, n : \left\|\mathbf{X}_{i,n/k}\right\|_\infty > 1\right\}.$$

We consider an identifiable (pseudo)-parametric statistical model $\mathcal{F}_k = \{\ell_{\mathbf{X}^*}(\theta; \mathbf{x}_{i,k}), \theta \in \Theta, \mathbf{x}_{i,k} \in \mathcal{A}, i \in \mathcal{N}_k\}$ whose (pseudo) log-likelihood, for $\mathbf{x} \in \mathcal{A}$, is given by

$$\ell_{\mathbf{X}^*}(\theta; \mathbf{x}) = \sum_{B \in \mathcal{P}} \log \mu\left(\theta; B, (\mathbf{x}_B, \mathbf{e}_{B^c})\right) \mathbb{I}_{\{\mathbf{x} \in \mathcal{A}_B\}} - \log \sum_{l=1}^{m} \mu\left(\theta; \{l\}, \mathbf{e}\right).$$

Conditions $(C1)$ and $(C2)$ of Section 3.1 imply the existence of the score function $\nabla_\theta \ell_{\mathbf{X}^*}(\theta; \mathbf{x})$ and the Hessian function $\nabla_\theta^2 \ell_{\mathbf{X}^*}(\theta; \mathbf{x})$ (see Section 6.5 for their expressions).

The maximum likelihood estimator $\hat{\theta}_k$ is defined by the following condition

$$\sum_{i \in \mathcal{N}_k} \nabla_\theta \ell_{\mathbf{X}^*}\left(\hat{\theta}_k; \mathbf{x}_{i,n/k}\right) = 0. \tag{3.6}$$

**Proposition 5** *Assume that* $(C1)$, $(C2)$ *and* $(C3)$ *hold. If, as* $n \to \infty$, $k \to \infty$ *such that* $k = o\left(n^{2\alpha/(1+2\alpha)}\right)$, *then*

$$\sqrt{k}(\hat{\theta}_k - \theta_0) \xrightarrow{d} \mathcal{N}\left(0, V_*^{-1}(\theta_0; \mathbf{e}) I_{\mathbf{X}^*}^{-1}(\theta_0)\right)$$

*where* $I_{\mathbf{X}^*}(\theta) = \mathbb{E}\left[\nabla_\theta \ell_{\mathbf{X}^*}(\theta; \mathbf{X}^*) \nabla_\theta' \ell_{\mathbf{X}^*}(\theta; \mathbf{X}^*)\right]$.

The proof is given in Section 6.4.

**Componentwise maxima with additional information on maxima occurrences**

Let $1 \leq k \leq n$ and $\lfloor y \rfloor$ denote the integer part of $y$. For $j = 1, \ldots, k$, we define $\mathbf{M}_{j,n/k}$ as the vector of componentwise maxima over the sub-sample $(\mathbf{Y}_i)_{i=1+(j-1)\lfloor n/k \rfloor, \ldots, j\lfloor n/k \rfloor}$, and $\mathbf{R}_{j,n/k}$ as the respective partition that gives which componentwise maxima occur simultaneously. We consider the identifiable (pseudo)-parametric statistical model $\mathcal{G}_k = \{\ell_{(\mathbf{M},\mathbf{R})}(\theta; (\mathbf{z}_{j,n/k}, \pi_{j,n/k})), \theta \in \Theta, \mathbf{z}_{i,n/k} \in \mathbb{R}_+^m, \pi_{j,n/k} \in \Pi, j = 1, \ldots, k\}$ whose (pseudo) log-likelihood is given by

$$\ell_{(\mathbf{M},\mathbf{R})}(\theta; (\mathbf{z}, \pi)) = -\sum_{l=1}^{m} z_l \mu(\theta; \{l\}, \mathbf{z}) + \sum_{B \in \pi} \log \mu(\theta; B, \mathbf{z}).$$

Conditions $(C1)$ and $(C2)$ of Section 3.1 imply the existence of the score function $\nabla_\theta \ell_{(\mathbf{M},\mathbf{R})}(\theta; (\mathbf{z}, \pi))$ and the Hessian function $\nabla_\theta^2 \ell_{(\mathbf{M},\mathbf{R})}(\theta; (\mathbf{z}, \pi))$ (see Section 6.5 for their expressions).

The maximum likelihood estimator (MLEm) satisfies the condition

$$\sum_{j=1}^{k} \nabla_\theta \ell_{(\mathbf{M},\mathbf{R})}\left(\bar{\theta}_k; (k\mathbf{m}_{j,n/k}/n, \mathbf{r}_{j,n/k})\right) = 0. \tag{3.7}$$

**Proposition 6** *Assume that $(C1)$ and $(C2)$ hold and that $n \to \infty$ and $k \to \infty$. If the following condition holds*

$$\sqrt{k}\mathbb{E}\left[\nabla_\theta \ell_{(\mathbf{M},\mathbf{R})}\left(\theta_0; (k\mathbf{M}_{j,n/k}/n, \mathbf{R}_{j,n/k})\right)\right] \to 0,$$

*then*

$$\sqrt{k}(\bar{\theta}_k - \theta_0) \xrightarrow{d} \mathcal{N}(0, I_{(\mathbf{M},\mathbf{R})}^{-1}(\theta_0))$$

*where* $I_{(\mathbf{M},\mathbf{R})}(\theta) \equiv \mathbb{E}\left[\nabla_\theta \ell_{(\mathbf{M},\mathbf{R})}(\theta; (\mathbf{M}, \mathbf{R})) \nabla_\theta' \ell_{(\mathbf{M},\mathbf{R})}(\theta; (\mathbf{M}, \mathbf{R}))\right].$

The proof uses usual arguments and is therefore omitted.

## 3.4 Discussion

**Monte-Carlo integration for $\mu(\theta; B, \mathbf{z})$**

If a Monte-Carlo integration is performed as the numerical integration method for computing $\mu(\theta; B, \mathbf{z})$ and if the number of simulations is not sufficiently large, the asymptotic distributions of the estimators may be modified. This consequence has already been observed with the simulated likelihood inference methods (see e.g. [21] and the references therein).

Assume that it is possible to write $\mu(\theta; B, \mathbf{z})$ as an expectation of a function $a$ with respect to a random variable $V$ (independent of $\theta$, $B$ and $\mathbf{z}$) such that

$$\mu(\theta; B, \mathbf{z}) = \mathbb{E}[a(V; \theta, B, \mathbf{z})]$$

and assume that $\nabla_\theta a(v; \theta, B, \mathbf{z})$ exists for any $v$, $\theta$, $B$ and $\mathbf{z}$, and that $\mathbb{E}[a(V; \theta, B, \mathbf{z})] < \infty$. For example, one possible way could be to choose $V$ as a univariate random variable with a unit Pareto distribution and let

$$a(v; \theta, B, \mathbf{z}) = v^{-|B|}\lambda(v^{-1}; \theta, B, \mathbf{z}) + v^{2+|B|}\lambda(v; \theta, B, \mathbf{z})$$

with $\lambda(v; \theta, B, \mathbf{z}) = \Pr(\mathbf{U}_{B^c} \leq \mathbf{z}_{B^c}v | \mathbf{U}_B = \mathbf{z}_B v) f_{\mathbf{U}_B}(\mathbf{z}_B v)$. For a unique sample of size $S$ of $V$, estimates of $\mu(\theta; B, \mathbf{z})$ are then given by

$$\mu_S(\theta; B, \mathbf{z}) = \frac{1}{S} \sum_{s=1}^{S} a(V_s; \theta, B, \mathbf{z})$$

(we use the same sample $(V_s)_{s=1,\ldots,S}$ for all estimates).

Let us consider for example the case of the maximum likelihood estimator for a sample of max-stable vectors as in Section 3.2. The simulated log-likelihood $\ell_{1S}$ has the same expression as $\ell_{1S}$ but $\mu$ is now replaced by $\mu_S$. The simulated maximum likelihood estimator (SMLE) $\hat{\theta}_{n,S}^{(1)}$ then satisfies

$$\sum_{i=1}^{n} \nabla_\theta \ell_{1S}(\hat{\theta}_{n,S}^{(1)}; \mathbf{z}_i) = 0. \tag{3.8}$$

**Proposition 7** *Assume that $n$ and $S$ tend to infinity. Under the regularity conditions of Proposition 4,*
  - *if $n/S$ tends to zero, then*

$$\sqrt{n}(\hat{\theta}_{n,S}^{(1)} - \theta_0) \xrightarrow{d} \mathcal{N}\left(0, I_1^{-1}(\theta_0)\right),$$

  - *if $n/S$ tends to infinity, then*

$$\sqrt{S}(\hat{\theta}_{n,S}^{(1)} - \theta_0) \xrightarrow{d} \mathcal{N}\left(0, I_1^{-1}(\theta_0)\,\Sigma(\theta_0)\,I_1^{-1}(\theta_0)\right)$$

*with $\Sigma(\theta_0) = \mathbb{V}\left(\mathbb{E}\left[\psi\left(V;\theta_0,\pi,\mathbf{Z}\right)|V\right]\right)$ where function $\psi$ is given in Section 6.6*
  - *if $n/S$ tends to a positive constant $\varphi < \infty$, then*

$$\sqrt{n}(\hat{\theta}_{n,S}^{(1)} - \theta_0) \xrightarrow{d} \mathcal{N}\left(0, I_1^{-1}(\theta_0)\left(I_1(\theta_0) + \varphi\Sigma(\theta_0)\right)I_1^{-1}(\theta_0)\right).$$

The proof of this proposition is given in the Complementary Material.

**Assumption about marginal components with unit Fréchet or unit Pareto distributions**

i) It has been assumed in Section 3.2 that the marginal distributions of $\mathbf{Z}$ are unit Fréchet. It is however possible to consider the case when the marginal distributions belong to the class of generalized extreme value (GEV) distributions. The GEV distributions are characterized by location, scale, and shape parameters (see e.g. [11]). Generally these parameters are described through parsimonious regression models which are functions of covariates to avoid computational issues that could arise for large numbers of parameters. In this way likelihood based inference allows simultaneous assessment of the parameters of the tail-dependence function as well as the location, scale, and shape parameters of the marginal distributions (see an example for spatial extremes with the pairwise likelihood in [39]).

ii) It has been assumed in Section 3.3 that the marginal distributions of $\mathbf{Y}$ are (asymptotically) unit Pareto distributions. When it is not the case, it is common to transform the margins to have unit Pareto ones. If $F_{Y_j}$ is known, the following transformation for $Y_j$ is considered: $(1 - F_{Y_j}(Y_j))^{-1}$. If not, transformations that used the data must be proposed. These transformations modify in general the asymptotic variance of the estimators (see e.g. [13]). Let us consider here an example of transformations that leads the marginal components to be asymptotically unit Pareto distributed.

We consider a sample of random vectors $(\mathbf{Y}_i)_{i=1,\ldots,n}$ distributed as $\mathbf{Y} = (Y_1,\ldots,Y_m)$ where $\mathbf{Y}$ is in the max-domain of attraction of an extreme value distribution $G$ whose margins are Fréchet distributed with unknown positive parameters $\alpha_j$ and such that $G\left(\mathbf{z}^{1/\alpha}\right) = G_*(\mathbf{z})$. For $1 \le k \le n$, we then define

$$\hat{\mathbf{X}}_{i,n/k} = \left(\frac{\mathbf{Y}_i}{\hat{\mathbf{U}}_{\mathbf{Y}}(n/k)}\right)^{\hat{\boldsymbol{\alpha}}_n} \vee \mathbf{e}$$

where $\hat{\boldsymbol{\alpha}}_n = (\hat{\alpha}_{1,n}, \ldots, \hat{\alpha}_{m,n})'$ and $\hat{\mathbf{U}}_{\mathbf{Y}}(n/k) = (\hat{U}_{Y_1}(n/k), \ldots, \hat{U}_{Y_m}(n/k))'$ with

$$\frac{1}{\hat{\alpha}_{j,n}} = \frac{1}{k}\sum_{i=0}^{k-1}\log\frac{Y_{(n-i)j}}{Y_{(n-k)j}}, \quad \hat{U}_{Y_j}(n/k) = Y_{(n-k)j} \quad \text{and} \quad Y_{(1)j} \leq Y_{(2)j} \leq \ldots \leq Y_{(n)j}.$$

The number of censored observations is given by the cardinality of the set

$$\hat{\mathcal{N}}_k = \{i = 1, \ldots, n : ||\hat{\mathbf{X}}_{i,n/k}||_\infty > 1\},$$

and the maximum likelihood estimator satisfies the condition

$$\sum_{i \in \hat{\mathcal{N}}_k} \nabla_\theta \ell_{\mathbf{X}^*}\left(\hat{\theta}_k; \hat{\mathbf{x}}_{i,n/k}\right) = 0.$$

This estimator is also asymptotically Gaussian as in Section 3.3, but its variance is modified due to the non-linear transformation that uses the data. Let $(\mathbf{W}_B(\mathbf{x}_B), \mathbf{x}_B \in [\mathbf{e}_B, \infty))_{B \in \mathcal{P}}$ be independent zero-mean Gaussian random fields with covariance functions

$$Cov\left(\mathbf{W}_B(\mathbf{x}_B), \mathbf{W}_B(\mathbf{y}_B)\right) = \nu_B\left((\mathbf{x}_B, \infty] \cap (\mathbf{y}_B, \infty]\right)$$

where

$$\nu_B\left((\mathbf{x}_B, \infty]\right) = \int_0^\infty \Pr\left(\mathbf{U}_B > \gamma \mathbf{x}_B, \mathbf{U}_{B^c} \leq \gamma \mathbf{e}_{B^c}\right) d\gamma.$$

Then define

$$W_j(x_j) = \sum_{B \in \mathcal{P}, B \cap \{j\} \neq \varnothing} \mathbf{W}_B\left((1, \ldots, 1, x_j, 1, \ldots, 1)\right)$$

and let

$$\begin{aligned}
\psi_j &= \mathbb{E}[\nabla^2_{\theta x_j}\ell_{\mathbf{X}^*}(\theta_0; \mathbf{X}^*) X_j^* \log X_j^* | X_j^* > 1] \\
\omega_j &= \mathbb{E}[\nabla^2_{\theta x_j}\ell_{\mathbf{X}^*}(\theta_0; \mathbf{X}^*) X_j^* | X_j^* > 1].
\end{aligned}$$

**Proposition 8** *Assume that (C1), (C2) and (C3) hold. If, as $n \to \infty$, $k \to \infty$ such that $k = o\left(n^{2\alpha/(1+2\alpha)}\right)$, then*

$$\sqrt{k}(\hat{\theta}_k - \theta_0) \xrightarrow{d} \left(V_*^{-1}(\theta_0; \mathbf{e})I_{\mathbf{X}^*}^{-1}(\theta_0)\right)\sum_{B \in \mathcal{P}}\int_{\mathcal{A}_B}\nabla_\theta \ell_{\mathbf{X}^*}(\theta_0; \mathbf{x})\,\mathbf{W}_B(d\mathbf{x}_B)$$

$$- \left(V_*^{-1}(\theta_0; \mathbf{e})I_{\mathbf{X}^*}^{-1}(\theta_0)\right)\left[\sum_{j=1}^m(\psi_j - \omega_j)\,W_j(1) + \sum_{j=1}^m\alpha_j\omega_j\int_1^\infty W_j(x_j^{\alpha_j})\frac{dx_j}{x_j}\right].$$

Note that the random variable $\left(V_*^{-1}(\theta_0; \mathbf{e})I_{\mathbf{X}^*}^{-1}(\theta_0)\right)\sum_{B \in \mathcal{P}}\int_{\mathcal{A}_B}\nabla_\theta\ell_{\mathbf{X}^*}(\theta_0; \mathbf{x})\,\mathbf{W}_B(d\mathbf{x}_B)$ has a centered Gaussian distribution with a variance equal to the asymptotic variance of the maximum likelihood estimator computed when the marginal distributions are unit Pareto, $V_*^{-1}(\theta_0; \mathbf{e})I_{\mathbf{X}^*}^{-1}(\theta_0)$. The proof of this proposition is given in the Complementary Material.

## 4 Simulation studies

We design a number of simulation studies to investigate the use of our likelihood based methods for inference on the parameters of high dimensional extreme value distributions. We used functions of the R package that we developed, HiDimMaxStable, to estimate the parameters.

## 4.1 Monte Carlo experiments for random samples from the Schlather max-stable process

We first consider random samples from the Schlather max-stable process (see [46]) with a Whittle Matérn correlation function given by

$$\rho(h) = \frac{2^{1-\nu}}{\Gamma(\nu)} \left(\frac{h}{c}\right)^{\nu} K_{\nu}\left(\frac{h}{c}\right)$$

where $c$ and $\nu$ are respectively called the range and the smoothness parameters of the correlation function, $\Gamma$ is the gamma function, $K_{\nu}$ is the modified Bessel function of the third kind with order $\nu$ and $h$ is the Euclidean distance between two locations.

For 100 simulation replicates, we randomly generate $m = 50$ (resp. 100) locations uniformly in the square $[0, 2] \times [0, 2]$. We then simulate $n = 40$ (resp. 20) max-stable process realizations under each model at the sampled $m$ locations using the SpatialExtremes R package ([42]). In this case the spectral random vector $\mathbf{U}$ of the multivariate max-stable distribution has a multivariate Gaussian distribution.

Since the intensity of the dependence between two locations is a function of the distance between these locations ($\rho$ is an isotropic correlation function), we decided to gather sites for the partition likelihood with a usual K-means algorithm based on the distance between the locations irrespectively of the values of the max-stables processes (see e.g. [35]). Note that we increased the number of clusters until the largest size of the clusters become smaller than or equal to 5.

Table 1: Means of the partition (MpcLE) and pairwise marginal (MpmLE) maximum likelihood estimates with their standard deviations (in brackets) and their relative efficiencies ($RE$) for the Whittle Matérn correlation function. The statistics are computed over 100 repetitions for each configuration of $m$, $n$ and $(c, \nu)$. Both estimates are computed with the same optimisation algorithm (Nelder-Mead), the same initial conditions (the true values), the same relative convergence tolerance, as well as the same maximum number of iterations.

| Whittle Matérn $m/n$ | | | | | | |
|---|---|---|---|---|---|---|
| | 100/20 | | | 50/40 | | |
| | $c$ | $\nu$ | $RE_c/RE_v$ | $c$ | $\nu$ | $RE_c/RE_v$ |
| True | 1 | 1 | | 1 | 1 | |
| MpcLE | 1.01(0.14) | 1.00(0.05) | | 1.01(0.15) | 1.00(0.06) | |
| MpmLE | 1.01(0.39) | 1.17(0.62) | 7.62/155 | 1.02(0.37) | 1.07(0.45) | 6.18/51 |
| | $c$ | $\nu$ | $RE_c/RE_v$ | $c$ | $\nu$ | $RE_c/RE_v$ |
| True | 1 | 2 | | 1 | 2 | |
| MpcLE | 1.00(0.15) | 2.00(0.17) | | 1.00(0.12) | 2.01(0.13) | |
| MpmLE | 1.06(0.51) | 3.23(4.02) | 10.5/560 | 1.08(0.42) | 2.62(2.54) | 13.7/447 |
| | $c$ | $\nu$ | $RE_c/RE_v$ | $c$ | $\nu$ | $RE_c/RE_v$ |
| True | 0.5 | 1 | | 0.5 | 1 | |
| MpcLE | 0.50(0.08) | 1.01(0.08) | | 0.50(0.09) | 1.01(0.10) | |
| MpmLE | 0.5(0.30) | 1.64(2.34) | 13.6/854 | 0.53(0.18) | 1.09(0.48) | 4.37/21 |

We compared partition composite likelihood inference to that from pairwise marginal composite likelihood. The maximum likelihood estimates of $(c, \nu)$ were computed from the 100 simulation replicates and were used to get the empirical standard deviations, the empirical mean square errors and the empirical relative efficiencies ($RE$) which is equal to the ratio of the mean square errors for the pairwise marginal composite likelihood inference method over the mean square errors for the partition composite likelihood inference method. The estimates, with their standard deviations (in brackets) and their relative efficiencies, are reported in Table 1.

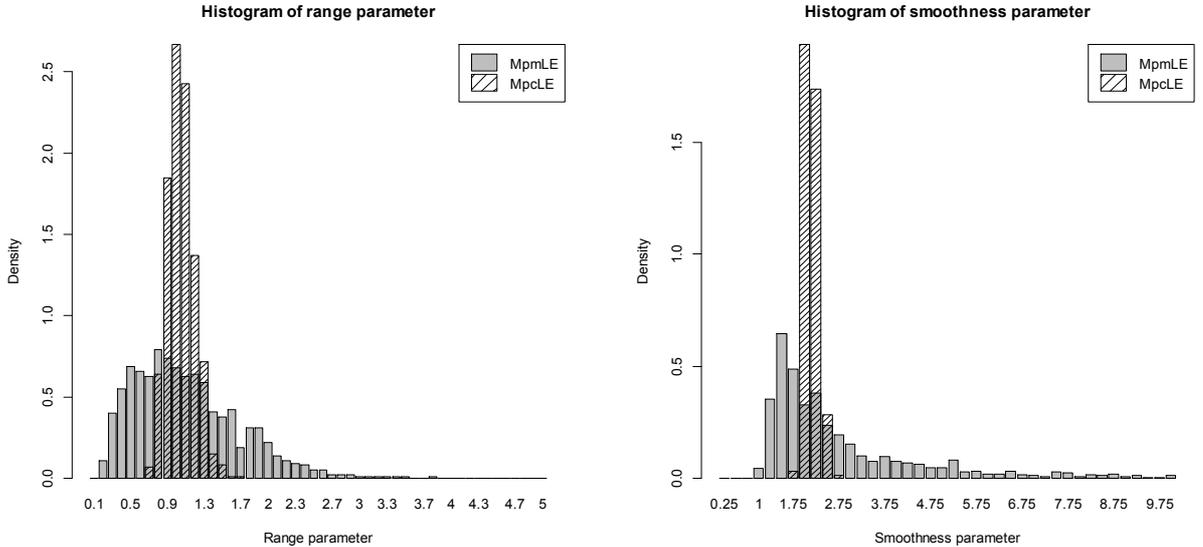

Figure 1: Histograms of the estimates of the range parameter ($c = 1$) and of the smoothness parameter ($\nu = 2$) in the set up $m = 100$, $n = 20$ for 1000 simulation replicates

We observe very small biases for the MpcLE of the range and the smoothness parameters. On the contrary the MpmLE of the smoothness parameters may have upward biases. The standard deviations of the MpcLE are always smaller than those of the MpmLE and the differences may be very large for some sets of parameters, e.g. $c = 1$ and $\nu = 2$. Histograms in Figure 1 show that the distributions of MpmLE are very spread out around their means and that they have heavier right tails. The pairwise marginal composite likelihood achieves more accurate parameter estimations in the set up $m = 50$ and $n = 40$ (more spatial observations but less temporal observations) while the partition composite likelihood gives roughly the same precision.

We also performed simulation in sets up with smaller values for the smoothness parameter ($\nu = 0.25$, $\nu = 0.5$, $\nu = 0.75$), a fixed value for $c$ ($= 1$), $m = 100$ and $n = 20$. We observed that the relative efficiencies of the smoothness parameter fall and are smaller than 10. Such results have already been described for the Brown-Resnick process in [24] where it is shown that the efficiency gains for the tirplewisecomposite likelihood are substantial only for smooth processes.

We conclude that there may be remarkable efficiency gains by using the partition composite likelihood inference when the smoothness parameter takes values larger than 1 (but with a longer computer time).

## 4.2 Monte Carlo experiments for (censored) random samples in the max-domain of attraction of a max-stable distribution with a Gaussian spectral structure

We now simulate random samples $(\mathbf{Y}_i)_{i=1,\dots,n}$ distributed as $\mathbf{Y} = (Y_1, \dots, Y_m)$ in the max-domain of attraction of a max-stable distribution with a Gaussian spectral structure. We consider the following model: $Y_j = U(x_j)/R$ for $j = 1, \dots, m$ where $R$ is a random variable with uniform distribution on $[0, 1]$, $U = \sqrt{2\pi} T$ where $T$ is a spatial stationary Gaussian process with unit variance and a Whittle Matérn correlation function as in the previous section, and $x_1, \dots, x_m$ are the site locations. We assume moreover that $R$ and $(U(x_1), \ldots, U(x_m))$ are independent. Note that it is the same type of model that has been studied in the case of a sample in the max-domain of attraction of the Brown–Resnick process in [53].

For 100 simulation replicates, we randomly generate $m = 20$ locations uniformly in the square $[0, 2] \times [0, 2]$. We then simulate $n = 250$ (resp. 1000) realizations of $\Gamma$ and $(U(x_1), \ldots, U(x_m))$ under each model at the sampled $m$ locations $(x_i)_{i=1,\ldots,m}$. The (censored) maximum likelihood estimates of $(c, \nu)$ with their empirical standard deviations were computed from these replicates. They are reported in Table 2.

Table 2: Means of the maximum likelihood estimates with their standard deviations (in brackets) for the Whittle Matérn correlation function. The statistics are computed over 100 repetitions for each configuration of $n$, $k/n$ and $(c, \nu)$.

| WM | | $c$ | | | | $\nu$ | | |
|---|---|---|---|---|---|---|---|---|
| $n = 250$ | 5% | 10% | 15% | 20% | 5% | 10% | 15% | 20% |
| $(1, 1)$ | 0.96(0.29) | 0.96(0.24) | 1.01(0.19) | 1.00(0.17) | 1.09(0.21) | 1.07(0.19) | 1.02(0.12) | 1.01(0.10) |
| $(1, 2)$ | 1.02(0.25) | 0.99(0.14) | 0.99(0.11) | 0.99(0.10) | 2.04(0.28) | 2.03(0.18) | 2.02(0.14) | 2.01(0.13) |
| $(0.5, 1)$ | 0.52(0.19) | 0.52(0.12) | 0.51(0.10) | 0.50(0.09) | 1.10(0.36) | 1.04(0.22) | 1.03(0.18) | 1.03(0.16) |
| $n = 1000$ | 5% | 10% | 15% | 20% | 5% | 10% | 15% | 20% |
| $(1, 1)$ | 1.00(0.17) | 0.99(0.10) | 0.99(0.09) | 0.98(0.08) | 1.00(0.09) | 1.00(0.06) | 1.00(0.06) | 1.00(0.05) |
| $(1, 2)$ | 1.00(0.11) | 1.00(0.07) | 0.99(0.06) | 0.99(0.05) | 2.02(0.13) | 2.01(0.08) | 2.01(0.07) | 2.01(0.06) |
| $(0.5, 1)$ | 0.49(0.08) | 0.50(0.05) | 0.50(0.05) | 0.50(0.04) | 1.03(0.15) | 1.00(0.09) | 1.00(0.07) | 1.00(0.06) |

We observe very small biases for the MLE with the Whittle Matérn correlation function. As expected, efficiency is improved when the numbers of observations $n$ increases and/or the ratio $k/n$ (the reciprocal of the threshold used to censor the observations) increases. From this Monte-Carlo experiment, we may conclude that our inference method performs very well for samples of moderate size generated from spatial processes. Note that we limited the number of locations to 20 because the R package mnormpow that we used for $\mu(B; \mathbf{z})$ (see section 6.2) is not able to provide computations of the moments of the multivariate normal distributions for dimension larger than 20.

## 4.3 Monte Carlo experiments for (censored) random samples in the max-domain of attraction of a clustered max-stable distribution

We now simulate random samples $(\mathbf{Y}_i)_{i=1,\ldots,n}$ distributed as $\mathbf{Y} = (Y_1, \ldots, Y_m)$ in the max-domain of attraction of an homogeneous clustered max-stable distribution as defined in Section 2.2. We assume that $\mathbf{Y} = \Gamma \mathbf{U} + \mathbf{E}$ where $\Gamma$ is a random variable with unit Pareto distribution independent of $\mathbf{U}$ and $\mathbf{E}$. For 100 simulation replicates, we randomly generate $\Gamma$, $\mathbf{U}$ and $\mathbf{E}$ with the following assumptions:

- $n = 2500$, $m = 100$ and $I = 3$ with $P_1 = \{1, \ldots, 50\}$, $P_2 = \{51, \ldots, 80\}$ and $P_3 = \{81, \ldots, 100\}$;

- $C_1$ is the Gumbel copula with generator $\psi_1(t) = \exp(-t^{1/\theta_1})$ and $\theta_1 = 1.7$, $C_2$ is the Clayton copula with generator $\psi_2(t) = (1 + u)^{-1/\theta_2}$ and $\theta_2 = 0.4$, $C_3$ is the Gumbel copula with parameter $\theta_3 = 1.2$;

- $F_1$ is the Log-normal distribution associated to the Gaussian distribution with standard error $\alpha_1 = 0.9$ and mean equal to $-\alpha_1^2/2$, $F_2$ is the Weibull distribution with shape parameter $\alpha_2$ equal to 1.5 and a scale parameter such that its expectation is equal to one, $F_3$ is the Fréchet distribution with shape parameter $\alpha_3$ equal to 1.7 and the scale parameter such that its expectation is equal to one.

- **E** is a vector of independent components with common Exponential distribution with mean equal to 10.

To identify the particular structure of **U**, consider the vector $\mathbf{Y}/\|\mathbf{Y}\|$ where $\|\mathbf{Y}\| = m^{-1} \sum_{j=1}^{m} Y_j$. Since $m$ is large, $\|\mathbf{E}\|$ is close to the mean of the $E_j$, i.e. 10. $\|\mathbf{U}\|$ has an expectation equal to one and we deduce that, when $\Gamma$ (or $\|\mathbf{Y}\|$) is large, $\mathbf{Y}/\|\mathbf{Y}\|$ is close to $\mathbf{U}/\|\mathbf{U}\|$. Therefore, keep vectors $\mathbf{Y}_i$ for which $\|\mathbf{Y}_i\|$ is larger than a given threshold and compute Kendall's tau coefficients for the components of the approximated spectral random vectors $(\mathbf{Y}_i/\|\mathbf{Y}_i\|)$. The heatmap of the matrix of Kendall's tau coefficients is plotted on Figure 2 for the threshold 20.

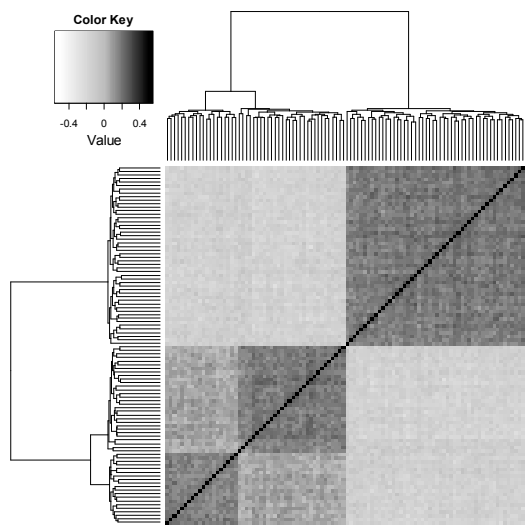

Figure 2: Heatmap of the matrix of Kendall's tau coefficients of the components of the approximated spectral random vectors $\mathbf{Y}_i/\|\mathbf{Y}_i\|$ for vectors $\mathbf{Y}_i$ satisfying $\|\mathbf{Y}_i\| > 20$.

We also give dendrograms to the left side and to the top after having reordered the rows and columns. A partition in two clusters appears clearly, and a partition in three clusters is also visible. Note that an artificial dependence is created between cluster 2 and cluster 3, because the $\|\mathbf{U}_i\|$ can't be approximated by 1 for sample with multivariate Archimedean copulas, and then the $\mathbf{U}_i/\|\mathbf{U}_i\|$ don't have the same distribution as **U**.

The statistical procedure we implemented is based on the likelihood for censored observations and follows two steps. Firstly we estimated separately the pairs $(\theta_i, \alpha_i)$ for each cluster ($i = 1, 2, 3$) of the partition. Secondly we estimated jointly all the parameters by using (2.4) for the log-likelihood with initial conditions given by estimates of the pairs. Every time we transformed the marginal distributions of the observations such that they became approximately unit Pareto distributions by using their respective order statistics and chose the threshold $t = 10$.

The results are presented in the following table.

Table 3. Means, standard errors and root-mean-square errors of parameter estimates of the copulas and of the marginal distributions for the first and the second step. The standard errors have been computed by simulations.

| | Cluster 1 (size=50) | | Cluster 2 (size=30) | | Cluster 3 (size=20) | |
|---|---|---|---|---|---|---|
| | Copula | Marg. dist. | Copula | Marg. dist. | Copula | Marg. dist. |
| | Gumbel | LogNormal | Clayton | Weibull | Gumbel | Fréchet |
| $(\theta_i, \alpha_i)$ | 1.7 | 0.9 | 0.4 | 1.5 | 1.2 | 1.7 |
| Means (step 1) | 1.691 | 0.905 | 0.422 | 1.489 | 1.218 | 1.705 |
| $s.e.$ (step 1) | 0.060 | 0.033 | 0.037 | 0.033 | 0.064 | 0.106 |
| RMSE (step 1) | 0.061 | 0.034 | 0.043 | 0.035 | 0.066 | 0.106 |
| Means (step 2) | 1.704 | 0.913 | 0.431 | 1.450 | 1.223 | 1.731 |
| $s.e.$ (step 2) | 0.041 | 0.022 | 0.035 | 0.031 | 0.024 | 0.038 |
| RMSE (step 2) | 0.041 | 0.026 | 0.047 | 0.058 | 0.033 | 0.049 |

We observe very small biases for the estimates of the parameters of the copulas and of the marginal distributions in the first step of the estimation procedure. The second step may increase slightly these biases since the whole distribution is now taken into consideration, but it mainly reduces significantly the standard errors of the estimators which leads to smaller root-mean-square errors (except for the parameter of the Clayton copula). This Monte-Carlo experiment shows that clustered max-stable distributions may be efficiently estimated and therefore that they can be used to model high-dimensional extreme events in practice.

## 5   Conclusion and discussion

We studied the problem of estimating the parameters of high dimensional extreme value distributions whose spectral random vectors have absolutely continuous distributions. These distributions appear naturally for the finite-dimensional distributions of the max-stable processes which are used to study extreme value phenomena for which spatial patterns can be discerned. For these distributions, we provided quasi-explicit analytical expressions of the full likelihoods for random samples with max-stable distributions and for random samples in the max-domain of attraction of a max-stable distribution. Monte Carlo experiments show that our likelihood-based methods of inference are efficient. It is important to underline that our inference methods may be used even for models for which the bivariate max-stable distributions are analytically unknown and for which the pairwise composite likelihood doesn't work. Finally we are convinced that the family of homogeneous clustered max-stable models, coupled with the very rich copula toolbox, could open new opportunities for the development of high dimensional extreme data analysis.

# 6 Appendix

## 6.1 Proof of Proposition 1

We have

$$
\begin{aligned}
V_*(\mathbf{z}) &= \mathbb{E}\left[\max_{j=1,\ldots,m} z_j^{-1} U_j^+\right] = \mathbb{E}\left[\left(\max_{j=1,\ldots,m} z_j^{-1} U_j\right)^+\right] = \int_0^\infty \Pr\left(\max_{j=1,\ldots,m} z_j^{-1} U_j > \gamma\right) d\gamma \\
&= \int_0^\infty \left[1 - \Pr\left(\max_{j=1,\ldots,m} z_j^{-1} U_j \le \gamma\right)\right] d\gamma = \int_0^\infty \left[1 - \Pr(\mathbf{U} \le \gamma \mathbf{z})\right] d\gamma.
\end{aligned}
$$

We deduce that

$$
\begin{aligned}
\frac{\partial^{|B|}}{\partial \mathbf{z}_B} V_*(\mathbf{z}) &= -\int_0^\infty \frac{\partial^{|B|}}{\partial \mathbf{z}_B} \Pr(\mathbf{U} \le \gamma \mathbf{z}) d\gamma \\
&= -\int_0^\infty \gamma^{|B|} \Pr\left(\mathbf{U}_{B^c} \le \mathbf{z}_{B^c}\gamma \mid \mathbf{U}_B = \mathbf{z}_B\gamma\right) f_{\mathbf{U}_B}(\mathbf{z}_B\gamma) \, d\gamma \\
&= -\mu\left(B; \mathbf{z}\right).
\end{aligned}
$$

Since $V_*(\mathbf{z})$ is an homogeneous function of order $-1$, Euler's homogeneous theorem implies that

$$
V_*(\mathbf{z}) = -\sum_{i=1}^m z_i \frac{\partial V_*(\mathbf{z})}{\partial z_i} = \sum_{i=1}^m z_i \mu\left(\{i\}; \mathbf{z}\right)
$$

and the result follows.

## 6.2 $\mu\left(B; \mathbf{z}\right)$ for the Gaussian and Log-normal models

A) Assume that $\mathbf{U}$ has a centered multivariate Gaussian distribution with covariance matrix $\Sigma$ such that $\mathbb{E}[U_j^+] = 1$, for $j = 1, \ldots, m$. Let us denote by $\Sigma_B$ the covariance matrix of $\mathbf{U}_B$ and $\Sigma_{B^c B}$ the covariance matrix between $\mathbf{U}_{B^c}$ and $\mathbf{U}_B$. Let $\|\mathbf{z}_B\|_{\Sigma_B^{-1}}^2 = \mathbf{z}_B' \Sigma_B^{-1} \mathbf{z}_B$, $\tilde{\mathbf{z}}_{B^c} = \left(\mathbf{z}_{B^c} - \Sigma_{B^c B} \Sigma_B^{-1} \mathbf{z}_B\right) / \|\mathbf{z}_B\|_{\Sigma_B^{-1}}$. Assume that $\mathbf{V}_{B^c|B} \sim \mathcal{N}\left(0, \Sigma_{B^c|B}\right)$ with $\Sigma_{B^c|B} = \Sigma_{B^c} - \Sigma_{B^c B} \Sigma_B^{-1} \Sigma_{BB^c}$ and $\Lambda \sim \mathcal{N}(0, 1)$ independent of $\mathbf{V}_{B^c|B}$. Then define $\mathbf{Y}$ as a random vector of dimension $|B^c| + 1$ in the following way: $\mathbf{Y}_{B^c} = \mathbf{V}_{B^c|B} + \tilde{\mathbf{z}}_{B^c}\Lambda$ and $Y_{\{|B^c|+1\}} = \Lambda$. It may be shown that

$$
\mu\left(B; \mathbf{z}\right) = \frac{1}{(2\pi)^{m/2}} \frac{\|\mathbf{z}_B\|_{\Sigma_B^{-1}}^{-(|B|+1)}}{(\det \Sigma_B)^{1/2} (\det \Sigma_\mathbf{Y})^{1/2}} \int_{-\infty}^0 \cdots \int_{-\infty}^0 |y_{|B^c|+1}|^{|B|} \exp\left(-\frac{1}{2}\mathbf{y}' \Sigma_Y^{-1} \mathbf{y}\right) d\mathbf{y}.
$$

These integrals are efficiently computed by using the approach developed by Genz (see [17], [18] and [20]).

Such an assumption about $\mathbf{U}$ holds for the (spatial) Schlather max-stable process (see [46]).

B) Assume that $\ln \mathbf{U} = \boldsymbol{\varepsilon} - \boldsymbol{\nu}$ where $\boldsymbol{\varepsilon}$ has a centered multivariate Gaussian distribution with covariance matrix $\Sigma$ and $\boldsymbol{\nu}$ is a vector such that $\nu_i = \mathbb{V}ar(\varepsilon_i)/2$. Let us denote by $\Sigma_B$ the covariance matrix of $\boldsymbol{\varepsilon}_B$ and by $\Sigma_{B^c B}$ the covariance matrix between $\boldsymbol{\varepsilon}_{B^c}$ and $\boldsymbol{\varepsilon}_B$. Let $\|\mathbf{z}_B\|_{\Sigma_B^{-1}}^2 = \mathbf{z}_B' \Sigma_B^{-1} \mathbf{z}_B$, $\mathbf{e} = (1, \ldots, 1)'$, $\langle \mathbf{e}_B | \log \mathbf{z}_B \rangle_{\Sigma_{|B}^{-1}} = \mathbf{e}_B' \Sigma_B^{-1} \mathbf{z}_B$, and define

$$
\log \tilde{\mathbf{z}}_B = \log \mathbf{z}_{B^c} + \boldsymbol{\nu}_{B^c} - \Sigma_{B^c B} \Sigma_B^{-1} \left(\log \mathbf{z}_B + \boldsymbol{\nu}_B\right) - \left(\mathbf{e}_{B^c} - \Sigma_{B^c B} \Sigma_B^{-1} \mathbf{e}_B\right) \|\mathbf{e}_B\|_{\Sigma_B^{-1}}^{-2} \left(\langle \mathbf{e}_B | \log \mathbf{z}_B \rangle_{\Sigma_B^{-1}} - 1\right)
$$

Assume that $\mathbf{V}_{B^c|B} \sim \mathcal{N}\left(0, \Sigma_{B^c|B}\right)$ with $\Sigma_{B^c|B} = \Sigma_B - \Sigma_{B^cB}\Sigma_B^{-1}\Sigma_{BB^c}$ and $\Lambda \sim \mathcal{N}(0, \|\mathbf{e}_B\|_{\Sigma_B^{-1}}^{-2})$ is independent of $\mathbf{V}_{B^c|B}$. Then define $\mathbf{Y}$ as a random vector of dimension $|B^c|$ in the following way: $\mathbf{Y}_{B^c} = \mathbf{V}_{B^c|B} - \left(\mathbf{e}_{B^c} - \Sigma_{B^cB}\Sigma_B^{-1}\mathbf{e}_B\right)\Lambda$. It may be shown that

$$
\begin{aligned}
\mu\left(B; \mathbf{z}\right) &= \frac{1}{(2\pi)^{(|B|-1)/2}} \frac{\|\mathbf{e}_B\|_{\Sigma_B^{-1}}^{-1}}{(\det \Sigma_B)^{1/2} \prod_{i=1}^{|B|} z_i} \Phi_{\Sigma_{\mathbf{Y}}}\left(\log \tilde{\mathbf{z}}_B\right) \\
&\quad \times \exp\left(\frac{1}{2}\|\mathbf{e}_B\|_{\Sigma_B^{-1}}^2 \left(\frac{\langle\mathbf{e}_B|\log \mathbf{z}_B + \boldsymbol{\nu}_B\rangle_{\Sigma_B^{-1}} - 1}{\|\mathbf{e}_B\|_{\Sigma_B^{-1}}^2}\right)^2 - \frac{1}{2}\|\log \mathbf{z}_B + \boldsymbol{\nu}_B\|_{\Sigma_B^{-1}}^2\right)
\end{aligned}
$$

where $\Phi_{\Sigma_{\mathbf{Y}}}\left(\log \tilde{\mathbf{z}}_B\right)$ may be efficiently computed by using the approach developed by Genz (see [17], [18] and [20]). Therefore $\mu\left(B; \mathbf{z}\right)$ may be written as a quantity proportional to a $|B^c|$-dimensional multivariate normal probability (see also [14] for an equivalent approach).

Such an assumption about $\mathbf{U}$ holds for the spatial Brown-Resnick max-stable process. The vector $\boldsymbol{\varepsilon}$ is extracted from a zero mean intrinsically stationary Gaussian process $\varepsilon$ with semivariogram function $\nu$ and satisfying $\varepsilon(0) = 0$ almost surely (see [29]). It also holds for the spatial Geometric Gaussian max-stable process. The vector $\boldsymbol{\varepsilon}$ is extracted in this case from a zero mean stationary Gaussian process.

### 6.3 Proof of Proposition 2

Let us begin by an intermediate lemma.

**Lemma 9** *Let $\varphi : [0, \infty) \to \mathbb{R}^+$ be such that $\int_0^\infty \varphi\left(\gamma^{-1}\right) d\gamma < \infty$. If $\Gamma$ is a random variable with a unit Pareto distribution, then*

$$
\lim_{n \to \infty} n\mathbb{E}\left[\varphi\left(n^{-1}\Gamma\right)\right] = \int_0^\infty \varphi\left(\gamma^{-1}\right) d\gamma.
$$

*Proof:* We have

$$
n\mathbb{E}\left[\varphi\left(n^{-1}\Gamma\right)\right] = n\int_1^\infty \varphi\left(n^{-1}\gamma\right)\gamma^{-2} d\gamma = \int_0^n \varphi\left(v^{-1}\right) dv \to \int_0^\infty \varphi\left(v^{-1}\right) dv. \quad \square
$$

If $Y = \Gamma U$ where $U$ is a random variable independent of $\Gamma$ such that $\mathbb{E}\left[|U|\right] < \infty$, then we deduce from the previous lemma that

$$
\lim_{n \to \infty} n\Pr\left(Y > yn\right) = y^{-1}\mathbb{E}\left[U^+\right]
$$

since $n\Pr\left(Y > yn\right) = n\mathbb{E}\left[\bar{F}_U\left(yn/\Gamma\right)\right]$ where $\bar{F}_U(u) = \Pr\left(U > u\right)$. $\blacksquare$

Let us now assume that $\tilde{\mathbf{Y}} = \left(\tilde{Y}_1, \ldots, \tilde{Y}_m\right)$ is such that $\tilde{Y}_j = \Gamma U_j$ for $j = 1, \ldots, m$ where $\Gamma$ is a random variable with unit Pareto distribution and independent of $\mathbf{U} = \left(U_1, \ldots, U_m\right)$. Then $\tilde{\mathbf{Y}}$ belongs to the max-domain of attraction of $G_*$. Indeed we have

$$
\log \Pr\left(\mathbf{M}_n \leq n\mathbf{z}\right) = n\log \Pr\left(\Gamma \max_{j=1,\ldots,m} z_j^{-1} U_j \leq n\right)
$$

and it follows by Lemma 9 that

$$\lim_{n\to\infty} \log \Pr\left(\mathbf{M}_n \leq n\mathbf{z}\right) = -\lim_{n\to\infty} n \Pr\left(\Gamma \max_{j=1,\ldots,m} z_j^{-1} U_j > n\right) = -\mathbb{E}\left[\max_{j=1,\ldots,m} z_j^{-1} U_j^+\right].$$

We may therefore assume without loss of generality that $\mathbf{X}_t = t^{-1}\tilde{\mathbf{Y}} \vee \mathbf{e}$ instead of $t^{-1}\mathbf{Y} \vee \mathbf{e}$ to establish the weak convergence of $\mathbf{X}_t^*$ to $\mathbf{X}^*$. We have, for $B \in \mathcal{P}$ and $\mathbf{x} \in \mathcal{A}_B$,

$$\Pr\left(\mathbf{X}_{B,t}^* \in (\mathbf{e}_B, \mathbf{x}_B] | \mathbf{X}_t^* \in \mathcal{A}_B\right) = \Pr\left(\mathbf{X}_{B,t} \leq \mathbf{x}_B | \mathbf{X}_{B,t} > \mathbf{e}_B, \mathbf{X}_{B^c,t} = \mathbf{e}_{B^c}\right)$$

and

$$
\begin{aligned}
&\lim_{t\to\infty} \Pr\left(\mathbf{X}_{B,t} \leq \mathbf{x}_B | \mathbf{X}_{B,t} > \mathbf{e}_B, \mathbf{X}_{B^c,t} = \mathbf{e}_{B^c}\right) \\
=\ &\lim_{t\to\infty} \frac{t\Pr\left(\mathbf{e}_B < \mathbf{X}_{B,t} \leq \mathbf{x}_B, \mathbf{X}_{B^c,t} = \mathbf{e}_{B^c}\right)}{t\Pr\left(\mathbf{X}_{B,t} > \mathbf{e}_B, \mathbf{X}_{B^c,t} = \mathbf{e}_{B^c}\right)} \\
=\ &\lim_{t\to\infty} \frac{t\Pr(t\mathbf{e}_B < \tilde{\mathbf{Y}}_{B,t} \leq t\mathbf{x}_B, \tilde{\mathbf{Y}}_{B^c,t} \leq t\mathbf{e}_{B^c})}{t\Pr(\tilde{\mathbf{Y}}_{B,t} > t\mathbf{e}_B, \tilde{\mathbf{Y}}_{B^c,t} \leq t\mathbf{e}_{B^c})} \\
=\ &\frac{\int_0^\infty \Pr\left(\gamma\mathbf{e}_B < \mathbf{U}_B \leq \gamma\mathbf{x}_B, \mathbf{U}_{B^c} \leq \gamma\mathbf{e}_{B^c}\right) d\gamma}{\int_0^\infty \Pr\left(\mathbf{U}_B > \gamma\mathbf{e}_B, \mathbf{U}_{B^c} \leq \gamma\mathbf{e}_{B^c}\right) d\gamma}
\end{aligned}
$$

by using again Lemma 9. By differentiating with respect to the components of $\mathbf{x}_B$, we derive that the conditional density of $\mathbf{X}^*$ given that $\mathbf{X}^* \in \mathcal{A}_B$ is

$$f_{\mathbf{X}^*(\mathcal{A}_B)}(\mathbf{x}_B) = \frac{\mu\left(B; (\mathbf{x}_B, \mathbf{e}_{B^c})\right)}{V_B^*(\mathbf{e})}.$$

## 6.4  Proof of Proposition 5

The maximum likelihood estimator $\hat{\theta}_k$ satisfies the first order condition (3.6). For any $B \in \mathcal{P}$ and $A_B \subset (\mathbf{e}_B, \infty] \in \mathcal{A}_B$, let us define the empirical measure $N_{B,k}$ by

$$N_{B,k}(A_B) = \sum_{i=1}^n \mathbb{I}\{\mathbf{X}_{i,B,n/k} \in A_B, \mathbf{X}_{i,B^c,n/k} = \mathbf{e}_{B^c}\}.$$

By using e.g. Proposition 2.1 in [12], $k^{-1} N_{B,k}$ converges in the vague topology on $M_+\left((\mathbf{e}_B, \infty]\right)$, the space of positive Radon measures on $(\mathbf{e}_B, \infty]$, and therefore we have for $\mathbf{x}_B \in (\mathbf{e}_B, \infty]$

$$\frac{1}{k} N_{B,k}\left((\mathbf{x}_B, \infty]\right) \xrightarrow{\Pr} \lim_{t\to\infty} t\Pr\left(\mathbf{X}_{B,t} > \mathbf{x}_B, \mathbf{X}_{B^c,t} = \mathbf{e}_{B^c}\right).$$

By Lemma 9, this limit is equal to

$$\nu_B\left((\mathbf{x}_B, \infty]\right) = \int_0^\infty \Pr\left(\mathbf{U}_B > \gamma\mathbf{x}_B, \mathbf{U}_{B^c} \leq \gamma\mathbf{e}_{B^c}\right) d\gamma.$$

By using Condition $(C3)$ and since $k = o\left(n^{2\alpha/(1+2\alpha)}\right)$, it holds that

$$\lim_{k\to\infty} \sqrt{k}\left(\nu_B\left((\mathbf{x}_B, \infty]\right) - \frac{n}{k}\Pr\left(\mathbf{X}_{B,n/k} > \mathbf{x}_B, \mathbf{X}_{B^c,n/k} = \mathbf{e}_{B^c}\right)\right) = 0.$$

By Proposition 3.1 in [12], there exist zero-mean Gaussian random fields $(\mathbf{W}_B(\mathbf{x}_B), \mathbf{x}_B \in (\mathbf{e}_B, \infty))$ with covariance functions

$$Cov(\mathbf{W}_B(\mathbf{x}_B), \mathbf{W}_B(\mathbf{y}_B)) = \nu_B((\mathbf{x}_B, \infty] \cap (\mathbf{y}_B, \infty])$$

such that the random fields

$$\mathbf{W}_{B,n}(\mathbf{x}_B) = \sqrt{k}\left(\frac{1}{k}N_{B,k}((\mathbf{x}_B, \infty]) - \nu_B((\mathbf{x}_B, \infty])\right), \quad \mathbf{x}_B \in (\mathbf{e}_B, \infty),$$

converge weakly to $\mathbf{W}_B$ in the space of cadlag functions defined on $(\mathbf{e}_B, \infty)$, $D((\mathbf{e}_B, \infty))$, equiped with Skorohod's topology:

$$\mathbf{W}_{B,n}(\cdot) \Longrightarrow \mathbf{W}_B(\cdot). \tag{6.9}$$

Moreover there exist versions of $\mathbf{W}_B$ which are sample continuous. Note that $\mathbf{W}_B$ is independent of $\mathbf{W}_{B'}$ for $B \neq B'$ because the sets $\mathcal{A}_B$ and $\mathcal{A}_{B'}$ are disjoint.

The first order condition is equivalent to

$$0 = \frac{1}{k}\sum_{B \in \mathcal{P}}\int_{\mathcal{A}_B}\nabla_\theta \ell_{\mathbf{X}^*}(\hat{\theta}_k; \mathbf{x}) N_{B,k}(d\mathbf{x}_B)$$

where

$$\nabla_\theta \ell_{\mathbf{X}^*}(\theta; \mathbf{x})\mathbb{I}_{\{\mathbf{x} \in \mathcal{A}_B\}} = \frac{\nabla_\theta \mu(\theta; B, (\mathbf{x}_B, \mathbf{e}_{B^c}))}{\mu(\theta; B, (\mathbf{x}_B, \mathbf{e}_{B^c}))} - \frac{\nabla_\theta V^*(\theta; \mathbf{e})}{V^*(\theta; \mathbf{e})}.$$

An expansion around $\theta_0$ gives

$$\begin{aligned}
0 &= \frac{1}{k}\sum_{B \in \mathcal{P}}\int_{\mathcal{A}_B}\nabla_\theta \ell_{\mathbf{X}^*}(\theta_0; \mathbf{x}) N_{B,k}(d\mathbf{x}_B) \\
&\quad + (\hat{\theta}_k - \theta_0)\frac{1}{k}\sum_{B \in \mathcal{P}}\int_{\mathcal{A}_B}\nabla_\theta^2 \ell_{\mathbf{X}^*}(\theta_0; \mathbf{x}) N_{B,k}(d\mathbf{x}_B) + O_{\Pr}\left((\hat{\theta}_k - \theta_0)^2\right).
\end{aligned}$$

First note that

$$\frac{1}{k}\sum_{B \in \mathcal{P}}\int_{\mathcal{A}_B}\nabla_\theta^2 \ell_{\mathbf{X}^*}(\theta_0; \mathbf{x}) N_{B,k}(d\mathbf{x}) \xrightarrow{\Pr} \sum_{B \in \mathcal{P}}\int_{\mathcal{A}_B}\nabla_\theta^2 \ell_{\mathbf{X}^*}(\theta_0; \mathbf{x}) \nu_B(d\mathbf{x}_B))$$

and that this limit is equal to

$$\begin{aligned}
&V_*(\theta_0; \mathbf{e})\sum_{B \in \mathcal{P}}p_B(\theta_0; \mathbf{e})\int_{\mathcal{A}_B}\nabla_\theta^2 \ell_{\mathbf{X}^*}(\theta_0; \mathbf{x})\frac{\mu(\theta_0; B, (\mathbf{x}_B, \mathbf{e}_{B^c}))}{V_B^*(\theta_0; e)}d\mathbf{x}_B \\
&= V_*(\theta_0; \mathbf{e})\mathbb{E}\left[\nabla_\theta^2 \ell_{\mathbf{X}^*}(\theta_0; \mathbf{X}^*)\right].
\end{aligned}$$

Now observe that

$$\sum_{B \in \mathcal{P}}\int_{\mathcal{A}_B}\nabla_\theta \ell_{\mathbf{X}^*}(\theta_0; \mathbf{x})\,d\nu_B((\mathbf{x}_B, \infty)) = \sum_{B \in \mathcal{P}}\int_{\mathcal{A}_B}\nabla_\theta \ell_{\mathbf{X}^*}(\theta_0; \mathbf{x})\mu(\theta_0; B, (\mathbf{x}_B, \mathbf{e}_{B^c}))\,d\mathbf{x}_B = 0$$

and derive by (6.9) that

$$\begin{aligned}
&\frac{1}{\sqrt{k}}\int_{\mathcal{A}_B}\sum_{B \in \mathcal{P}}\nabla_\theta \ell_{\mathbf{X}^*}(\theta_0; \mathbf{x}) N_{B,k}(d\mathbf{x}_B) \\
&\xrightarrow{d} \mathcal{N}\left(0, \int_{\mathcal{A}_B}\sum_{B \in \mathcal{P}}\nabla_\theta \ell_{\mathbf{X}^*}(\theta_0; \mathbf{x})\nabla_\theta \ell_{\mathbf{X}^*}(\theta_0; \mathbf{x})'\mu(\theta_0; B, (\mathbf{x}_B, \mathbf{e}_{B^c}))\,d\mathbf{x}_B\right).
\end{aligned}$$

The asymptotic variance may be rewritten as

$$V_*(\theta; \mathbf{e}) \mathbb{E}\left[\nabla_\theta \ell_{\mathbf{X}^*}(\theta; \mathbf{X}^*) \nabla'_\theta \ell_{\mathbf{X}^*}(\theta; \mathbf{X}^*)\right]$$

by using Proposition 2. Since

$$I_{\mathbf{X}^*}(\theta_0) = \mathbb{E}\left[\nabla_\theta \ell_{\mathbf{X}^*}(\theta; \mathbf{X}^*) \nabla'_\theta \ell_{\mathbf{X}^*}(\theta; \mathbf{X}^*)\right] = -\mathbb{E}\left[\nabla^2_\theta \ell_{\mathbf{X}^*}(\theta_0; \mathbf{X}^*)\right],$$

we finally have

$$\sqrt{k}(\hat{\theta}_k - \theta_0) \xrightarrow{d} \mathcal{N}\left(0, V_*^{-1}(\theta_0; \mathbf{e}) I_{\mathbf{X}^*}^{-1}(\theta_0)\right).$$

## 6.5 Score and Hessian functions

### 6.5.1 Likelihoods for the max-stable distribution

Let $\chi(\theta; \pi, \mathbf{z}) = \prod_{B \in \pi} \mu(\theta; B, \mathbf{z})$ and $\delta(\theta; \mathbf{z}) = \sum_{\pi \in \Pi} \chi(\theta; \pi, \mathbf{z})$. The score functions are given by

$$
\begin{aligned}
\nabla_\theta \ell_1(\theta; \mathbf{z}) &= -\sum_{i=1}^m z_i \nabla_\theta \mu(\theta; \{i\}, \mathbf{z}) + \sum_{\pi \in \Pi} \frac{\chi(\theta; \pi, \mathbf{z})}{\sum_{\pi' \in \Pi} \chi(\theta; \pi', \mathbf{z})} \sum_{B \in \pi} \nabla_\theta \log \mu(\theta; B, \mathbf{z}), \\
\nabla_\theta \ell_2(\theta; \mathbf{z}) &= \sum_{B \in \pi} |B| \nabla_\theta \ell_1(\theta; \mathbf{z}_B), \\
\nabla_\theta \ell_3(\theta; \mathbf{z}) &= \sum_{i<j} \nabla_\theta \ell_1(\theta; \mathbf{z}_{\{i,j\}}),
\end{aligned}
$$

and their respective Hessian functions are

$$
\begin{aligned}
&\nabla^2_\theta \ell_1(\theta; \mathbf{z}) \\
={}& -\sum_{i=1}^m z_i \nabla^2_\theta \mu(\theta; \{i\}, \mathbf{z}) + \sum_{\pi \in \Pi} \frac{\chi(\theta; \pi, \mathbf{z})}{\sum_{\pi' \in \Pi} \chi(\theta; \pi', \mathbf{z})} \sum_{B \in \pi} \nabla^2_\theta \log \mu(\theta; B, \mathbf{z}) \\
&+ \sum_{\pi \in \Pi} \frac{\chi(\theta; \pi, \mathbf{z})}{\sum_{\pi' \in \Pi} \chi(\theta; \pi', \mathbf{z})} \sum_{B \in \pi} \nabla_\theta \log \mu(\theta; B, \mathbf{z}) \sum_{B' \in \pi} \nabla_\theta \log \mu(\theta; B', \mathbf{z})' \\
&- \left(\sum_{\pi \in \Pi} \frac{\chi(\theta; \pi, \mathbf{z})}{\sum_{\pi' \in \Pi} \chi(\theta; \pi', \mathbf{z})} \sum_{B \in \pi} \nabla_\theta \log \mu(\theta; B, \mathbf{z})\right) \left(\sum_{\pi \in \Pi} \frac{\chi(\theta; \pi, \mathbf{z})}{\sum_{\pi' \in \Pi} \chi(\theta; \pi', \mathbf{z})} \sum_{B \in \pi} \nabla_\theta \log \mu(\theta; B, \mathbf{z})\right)'
\end{aligned}
$$

and

$$\nabla^2_\theta \ell_2(\theta; \mathbf{z}) = \sum_{B \in \pi} |B| \nabla_\theta \ell_1(\theta; \mathbf{z}_B), \quad \nabla^2_\theta \ell_3(\theta; \mathbf{z}) = \sum_{i<j} \nabla^2_\theta \ell_1(\theta; \mathbf{z}_{\{i,j\}}).$$

### 6.5.2 Likelihood for the vector of exceedances with censored components

The score function is given by

$$\nabla_\theta \ell_{\mathbf{X}^*}(\theta; \mathbf{x}) = \sum_{B \in \mathcal{P}} \frac{\nabla_\theta \mu(\theta; B, (\mathbf{x}_B, \mathbf{e}_{B^c}))}{\mu(\theta; B, (\mathbf{x}_B, \mathbf{e}_{B^c}))} \mathbb{I}_{\{\mathbf{x} \in \mathcal{A}_B\}} - \frac{\sum_{i=1}^m \nabla_\theta \mu(\theta; \{i\}, \mathbf{e})}{\sum_{i=1}^m \mu(\theta; \{i\}, \mathbf{e})}$$

and its Hessian function by

$$
\begin{aligned}
&\nabla_\theta^2 \ell_{\mathbf{X}^*}(\theta; \mathbf{x}) \\
&= \sum_{B \in \mathcal{P}} \left( \frac{\nabla_\theta^2 \mu(\theta; B, (\mathbf{x}_B, \mathbf{e}_{B^c}))}{\mu(\theta; B, (\mathbf{x}_B, \mathbf{e}_{B^c}))} - \frac{\nabla_\theta \mu(\theta; B, (\mathbf{x}_B, \mathbf{e}_{B^c})) \nabla_\theta' \mu(\theta; B, (\mathbf{x}_B, \mathbf{e}_{B^c}))}{\mu^2(\theta; B, (x_B, e_{B^c}))} \right) \mathbb{I}_{\{\mathbf{x} \in \mathcal{A}_B\}} \\
&\quad - \left( \frac{\sum_{i=1}^m \nabla_\theta^2 \mu(\theta; \{i\}, \mathbf{e})}{\sum_{i=1}^m \mu(\theta; \{i\}, \mathbf{e})} - \frac{\sum_{i=1}^m \nabla_\theta \mu(\theta; \{i\}, \mathbf{e}) \sum_{j=1}^m \nabla_\theta' \mu(\theta; \{j\}, \mathbf{e})}{\left( \sum_{i=1}^m \mu(\theta; \{i\}, \mathbf{e}) \right)^2} \right).
\end{aligned}
$$

### 6.5.3 Likelihood for the componentwise maxima with maxima occurrences

The score function is given by

$$
\nabla_\theta \ell_{(\mathbf{M}, \mathbf{R})}(\theta; (\mathbf{z}, \pi)) = - \sum_{l=1}^m z_l \nabla_\theta \mu(\theta; \{l\}, \mathbf{z}) + \sum_{B \in \pi} \frac{\nabla_\theta \mu(\theta; B, \mathbf{z})}{\mu(\theta; B, \mathbf{z})}
$$

and its Hessian function by

$$
\begin{aligned}
&\nabla_\theta^2 \ell_{(\mathbf{M}, \mathbf{R})}(\theta; (\mathbf{z}, \pi)) \\
&= - \sum_{l=1}^m z_l \nabla_\theta^2 \mu(\theta; \{l\}, \mathbf{z}) + \sum_{B \in \pi} \left( \frac{\nabla_\theta^2 \mu(\theta; B, \mathbf{z})}{\mu(\theta; B, \mathbf{z})} - \frac{\nabla_\theta \mu(\theta; B, \mathbf{z}) \nabla_\theta' \mu(\theta; B, \mathbf{z})}{(\mu(\theta; B, \mathbf{z}))^2} \right).
\end{aligned}
$$

## 6.6 Function $\psi$

Function $\psi$ is given by

$$
\psi(v; \theta, \pi, \mathbf{z}) = \frac{1}{\delta(\theta; \mathbf{z})} \sum_{\pi \in \Pi} \chi(\theta; \pi, \mathbf{z}) \sum_{j=1}^3 \phi_j(v; \theta, \pi, \mathbf{z}) - \sum_{l=1}^m z_i \nabla_\theta a(v; \theta, \{l\}, \mathbf{z})
$$

where functions $\chi$ and $\delta$ are given in Section 6.5.1 and

$$
\begin{aligned}
\phi_1(v; \theta, \pi, \mathbf{z}) &= -\frac{\nabla_\theta \delta(\theta; \mathbf{z})}{\delta(\theta; \mathbf{z})} \sum_{B \in \pi} \frac{1}{\mu(\theta; B, \mathbf{z})} a(v; \theta, B, \mathbf{z}) \\
\phi_2(v; \theta, \pi, \mathbf{z}) &= \left( \sum_{B' \in \pi} \nabla_\theta \log \mu(\theta; B', \mathbf{z}) \right) \sum_{B \in \pi} \frac{1}{\mu(\theta; B, \mathbf{z})} a(v; \theta, B, \mathbf{z}) \\
\phi_3(v; \theta, \pi, \mathbf{z}) &= \sum_{B \in \pi} \frac{1}{\mu(\theta; B, \mathbf{z})} \left[ \nabla_\theta a(v; \theta, B, \mathbf{z}) - \nabla_\theta \log \mu(\theta; B, \mathbf{z}) a(v; \theta, B, \mathbf{z}) \right].
\end{aligned}
$$

# References

[1] Bacro, J.N. and Gaetan, C. (2013) Estimation of spatial max-stable models using threshold exceedances. arXiv:1205.1107.

[2] Bernard, E., Naveau, P., Vrac, M. and Mestre, O. (2013) Clustering of maxima: Spatial dependencies among heavy rainfall in france. *Journal of Climate*, to appear.

[3] Besag, J. (1974) Spatial interaction and the statistical analysis of lattice systems (with discussion). *J. R. Statist. Soc. B*, **36**, 192–236.

[4] Brown, B. and Resnick, S. (1977) Extremes values of independent stochastic processes. *Journal of Applied Probability*, **14**, 732–739.

[5] Castruccio, S., Huser, R. and Genton, M.G. (2014) High-order composite likelihood inference for multivariate or spatial extremes, arXiv:1411.0086v1.

[6] Cooley, D., Naveau, P. and Poncet, P. (2006) Variograms for spatial max-stable random fields. Dependence in Probability and Statistics, Lecture Notes In Statistics, 187, 373-320 90.

[7] Cox, D. and Reid, N. (2004) A note on pseudolikelihood constructed from marginal densities. *Biometrika*, **91**, 729–737.

[8] Davison, A.C. (2003) *Statistical Models*. New York: Cambridge University Press.

[9] Davison, A., Padoan, S. and Ribatet, M. (2012) Statistical modelling of spatial extremes. *Statistical Science*, **27**, 161–186.

[10] de Haan, L. (1984) A spectral representation for max-stable processes. *Annals of Probability*, **12**, 1194–1204.

[11] de Haan, L. and Ferreira, A. (2006) *Extreme Value Theory: an Introduction*. Springer Science Business Media, LLC, New York.

[12] de Haan, L. and Resnick, S.I. (1993) Estimating the limit distribution of multivariate extremes. *Communications in Statistics. Stochastic Models*, **9**, 275–309.

[13] Einmahl, J.H., Krajina, A. and Segers, J. (2012) An M-estimator for tail dependence in arbitrary dimensions. *The Annals of Statistics*, **40**, 1764–1793.

[14] Engelke, S., Malinowski, A., Kabluchko, Z. and Schlather, M. (2014) Estimation of Hüsler–Reiss distributions and Brown–Resnick processes. *J. R. Statist. Soc. B*, to appear.

[15] Galambos, J. (1975). Order statistics of samples from multivariate distributions. *Journal of the American Statistical Association*, **70**, 674–680.

[16] Genton, M.G., Ma, Y. and Sang, H. (2011) On the likelihood function of Gaussian max-stable processes. *Biometrika*, **98**, 481–488.

[17] Genz, A. (1992) Numerical computation of multivariate normal probabilities. *J. Computational and Graphical Statist.*, **1**, 141–149.

[18] Genz, A. (1993) Comparison of methods for the computation of multivariate normal probabilities. *Computing Science and Statistics*, **25**, 400–405.

[19] Genz, A.: Fortran code available at http://www.math.wsu.edu/math/faculty/genz/software/fort77/mvr

[20] Genz, A. and Bretz, F. (2009) *Computation of Multivariate Normal and t Probabilities*. Lecture Notes in Statistics **195**, Springer-Verlag, New York.

[21] Gouriéroux, C. and Monfort, A. (1996) *Simulation-based Econometric Methods*. Oxford University Press.

[22] Gumbel, E.J. (1960) Distributions des valeurs extrêmes en plusieurs dimensions. *Publication de l'ISUP*, **9**, 171–173.

[23] Hofert, M., Mächler, M. and McNeil, A.J. (2012) Likelihood inference for Archimedean copulas in high dimensions under known margins. *Journal of Multivariate Analysis*, **110**, 133–150.

[24] Huser, R. and Davison, A.C. (2013) Composite likelihood estimation for the Brown–Resnick process. *Biometrika*, **100**, 511–518.

[25] Huser, R., Davison, A.C. and Genton, M.G. (2014) A comparative study of likelihood estimators for multivariate extremes, arXiv:1411.3448v1.

[26] Hüsler, J. and Reiss, R.D. (1989) Maxima of normal random vectors: between independence and complete dependence. *Statistics and Probability Letters*, **9**, 283–286.

[27] Jeon, S. and Smith, R.L. (2012) Dependence structure of spatial extremes using threshold approach. arXiv:1209.6344.

[28] Joe, H., Li, H. and Nikoloulopoulos, A.K. (2010) Tail dependence functions and vine copulas. *Journal of Multivariate Analysis*, **101**, 252–270.

[29] Kabluchko, Z., Schlather, M. and de Haan, L. (2009) Stationary max-stable fields associated to negative definite functions. *Ann. Prob.*, **37**, 2042–2065.

[30] Klüppelberg, C., Kuhn, G. and Peng, L. (2008). Semi-parametric models for the multivariate tail dependence function - the asymptotically dependent. *Scandinavian Journal of Statistics*, **35**, 701–718.

[31] Kiriliouk, A. , Segers, J., Einmahl J. and Krajina, A. (2014). Estimation of tail dependence in high-dimensional and spatial models.

[32] Li, H. and Wu, P. (2013). Extremal dependence of copulas: A tail density approach. *Journal of Multivariate Analysis*, **114**, 99–111.

[33] Lindsay, B. (1988) Composite likelihood methods. *Contemp. Math.*, **80**, 221–239.

[34] Marshall, A.W. and Olkin, I. (1967) A multivariate exponential distribution. *Journal of the American Statistical Association*, **62**, 30–44.

[35] MacQueen, J.B. (1967). Some methods for classification and analysis of multivariate observations. Proceedings of 5th Berkeley Symposium on Mathematical Statistics and Probability 1. University of California Press., 281–297.

[36] McNeil, A.J., Frey, R. and Embrechts, P. (2005) *Quantitative Risk Management: Concepts, Techniques, Tools*. Princeton University Press.

[37] Nikoloulopoulos, A.K., Joe, H. and Li, H. (2009) Extreme value properties of multivariate t copulas. *Extremes,* **12**, 129–148.

[38] Optiz, T. (2012) Extremal t processes: Elliptical domain of attraction and a spectral representation. Working paper, available at arXiv:1207.2296.

[39] Padoan, S.A., Ribatet, M. and Sisson, S. (2010) Likelihood-based inference for max-stable processes. *Journal of the American Statistical Association*, **105**, 263–277.

[40] Penrose, M.D. (1992) Semi-min-stable processes. *Annals of Probability*, **20**, 1450-1463.

[41] Resnick, S. (1987) *Extreme values, Regular variation, and Point Processes.* Springer, New York.

[42] Ribatet, M. (2011). SpatialExtremes: modelling spatial extremes. R Package Version 1.8-0. URL http://CRAN.R-project.org/package=SpatialExtremes.

[43] Ribatet, M. (2013) Spatial extremes: Max-stable processes at work. *Journal of Société Française de Statistique*, **154**, 156–177.

[44] Rootzén, H. and Tajvidi, N. (2006) Multivariate generalized Pareto distributions. *Bernoulli*, **12**(5), 917–930

[45] Sang, H. and Genton, M.G. (2014) Tapered composite likelihood for spatial max-stable models. *Spatial Statistics*, **8**, 86–103.

[46] Schlather, M. (2002) Models for stationary max-stable random fields. *Extremes*, **5**, 33–44.

[47] Smith, R.L. (1990) Max-stable processes and spatial extremes. Preprint, University of Surrey, Surrey.

[48] Stephenson, A.G. and Tawn, J.A. (2005) Exploiting occurrence times in likelihood inference for componentwise maxima. *Biometrika*, **92**, 213–27.

[49] Varin, C. (2007) On composite marginal likelihoods. *Adv Statist. Anal.*, **92**, 1–28.

[50] Varin, C. and Vidoni, P. (2005) Pairwise likelihood inference and model selection. *Biometrika*, **92**, 519–528.

[51] Wadsworth, J. L. (2014). On the occurrence times of componentwise maxima and bias in likelihood inference for multivariate max-stable distributions.

[52] Wadsworth, J.L. and Tawn, J.A. (2012) Dependence modelling for spatial extremes. *Biometrika*, **99**, 253–272.

[53] Wadsworth, J.L. and Tawn, J.A. (2014) Efficient inference for spatial extreme value processes associated to log-Gaussian random functions. To appear in *Biometrika*.

# 7 Complementary material

## 7.1 Proof of Proposition 7

The SMLE, $\hat{\theta}_{n,S}^{(1)}$, is solution of

$$\frac{1}{n^{1/2}} \sum_{i=1}^{n} \nabla_\theta \ell_{1S}(\hat{\theta}_{n,S}^{(1)}; \mathbf{z}_i) = 0$$

where $\ell_{1S}(\theta; \mathbf{z}) = -\sum_{l=1}^{m} z_l \mu_S(\theta; \{l\}, \mathbf{z}) + \log(\delta_S(\theta; \mathbf{z}))$ with

$$\mu_S(\theta; B, \mathbf{z}) = \frac{1}{S} \sum_{s=1}^{S} a(V_s; \theta, B, \mathbf{z}), \quad \delta_S(\theta; \mathbf{z}) = \sum_{\pi \in \Pi} \chi_S(\theta; \pi, \mathbf{z}), \quad \chi_S(\theta; \pi, \mathbf{z}) = \prod_{B \in \pi} \mu_S(\theta; B, \mathbf{z}).$$

An expansion around $\theta_0$ gives

$$0 = \frac{1}{n^{1/2}} \sum_{i=1}^{n} \nabla_\theta \ell_{1S}(\theta_0; \mathbf{z}_i) + \left( \frac{1}{n} \sum_{i=1}^{n} \nabla_\theta^2 \ell_{1S}(\theta_0; \mathbf{z}_i) \right) n^{1/2}(\hat{\theta}_{n,S}^{(1)} - \theta_0) + O_{\mathrm{Pr}}\left( n^{1/2}(\hat{\theta}_{n,S}^{(1)} - \theta_0)^2 \right).$$

Let us begin by studying the difference

$$\frac{1}{n^{1/2}} \sum_{i=1}^{n} \nabla_\theta \ell_{1S}(\theta_0; \mathbf{z}_i) - \frac{1}{n^{1/2}} \sum_{i=1}^{n} \nabla_\theta \ell_1(\theta_0; \mathbf{z}_i).$$

Recall that

$$\nabla_\theta \ell_{1S}(\theta_0; \mathbf{z}) = -\sum_{l=1}^{m} z_l \nabla_\theta \mu_S(\theta_0; \{l\}, \mathbf{z}) + \frac{\nabla_\theta \delta_S(\theta_0; \mathbf{z})}{\delta_S(\theta_0; \mathbf{z})}$$

with

$$\nabla_\theta \delta_S(\theta_0; \mathbf{z}) = \sum_{\pi \in \Pi} \chi_S(\theta_0; \pi, \mathbf{z}) \sum_{B \in \pi} \nabla_\theta \log \mu_S(\theta_0; B, \mathbf{z}).$$

First it holds that

$$\nabla_\theta \mu_S(\theta_0; \{l\}, \mathbf{z}) - \nabla_\theta \mu(\theta_0; \{l\}, \mathbf{z}) = \int \nabla_\theta a(v; \theta_0, \{l\}, \mathbf{z}) \left( (dF_{V,S}(v) - dF_V(v)) \right)$$

where $F_{V,S}(v) = S^{-1} \sum_{s=1}^{S} \mathbb{I}_{\{V_s \leq v\}}$. Second,

$$\frac{\nabla_\theta \delta_S(\theta_0; \mathbf{z})}{\delta_S(\theta_0; \mathbf{z})} - \frac{\nabla_\theta \delta(\theta_0; \mathbf{z})}{\delta(\theta_0; \mathbf{z})}$$
$$= \frac{1}{\delta_S(\theta_0; \mathbf{z})} \left[ (\nabla_\theta \delta_S(\theta_0; \mathbf{z}) - \nabla_\theta \delta(\theta_0; \mathbf{z})) - \frac{\nabla_\theta \delta(\theta_0; \mathbf{z})}{\delta(\theta_0; \mathbf{z})} (\delta_S(\theta_0; \mathbf{z}) - \delta(\theta_0; \mathbf{z})) \right]$$

By the law of large numbers, $\mu_S(\theta_0; B, \mathbf{z}) \to_{\mathrm{Pr}} \mu(\theta_0; B, \mathbf{z})$ as $S \to \infty$. Since $\mathbb{E}\left[ a^2(V; \theta, B, \mathbf{z}) \right] < \infty$ for any $\mathbf{z} \in \mathbb{R}_+^m$, expansions around $\mu(\theta_0; B, \mathbf{z})$ give

$$\begin{aligned}
\delta_S(\theta_0; \mathbf{z}) - \delta(\theta_0; \mathbf{z}) &= \sum_{\pi \in \Pi} (\chi_S(\theta; \pi, \mathbf{z}) - \chi(\theta; \pi, \mathbf{z})) \\
&= \sum_{\pi \in \Pi} \chi(\theta_0; \pi, \mathbf{z}) \sum_{B \in \pi} \left( \frac{\mu_S(\theta_0; B, \mathbf{z}) - \mu(\theta_0; B, \mathbf{z})}{\mu(\theta_0; B, \mathbf{z})} \right) + O_{\mathrm{Pr}}\left( S^{-1} \right)
\end{aligned}$$

and

$$\begin{aligned}
&\nabla_\theta \delta_S(\theta_0; \mathbf{z}) - \nabla_\theta \delta(\theta_0; \mathbf{z}) \\
&= \sum_{\pi \in \Pi} \chi(\theta_0; \pi, \mathbf{z}) \sum_{B \in \pi} (\nabla_\theta \log \mu_S(\theta_0; B, \mathbf{z}) - \nabla_\theta \log \mu(\theta_0; B, \mathbf{z})) \\
&\quad + \sum_{\pi \in \Pi} \left( \sum_{B' \in \pi} \nabla_\theta \log \mu_S(\theta_0; B', \mathbf{z}) \right) (\chi(\theta_0; \pi, \mathbf{z}) - \chi_S(\theta_0; \pi, \mathbf{z})) \\
&= \sum_{\pi \in \Pi} \chi(\theta_0; \pi, \mathbf{z}) \sum_{B \in \pi} (\nabla_\theta \log \mu_S(\theta_0; B, \mathbf{z}) - \nabla_\theta \log \mu(\theta_0; B, \mathbf{z})) \\
&\quad + \sum_{\pi \in \Pi} \left( \sum_{B' \in \pi} \nabla_\theta \log \mu(\theta_0; B', \mathbf{z}) \right) \chi(\theta_0; \pi, \mathbf{z}) \sum_{B \in \pi} \left( \frac{\mu_S(\theta_0; B, \mathbf{z}) - \mu(\theta_0; B, \mathbf{z})}{\mu(\theta_0; B, \mathbf{z})} \right) + O_{\Pr}(S^{-1}).
\end{aligned}$$

It follows that

$$\begin{aligned}
&\frac{\nabla_\theta \delta_S(\theta_0; \mathbf{z})}{\delta_S(\theta_0; \mathbf{z})} - \frac{\nabla_\theta \delta(\theta_0; \mathbf{z})}{\delta(\theta_0; \mathbf{z})} \\
&= -\frac{\nabla_\theta \delta(\theta_0; \mathbf{z})}{\delta^2(\theta_0; \mathbf{z})} \sum_{\pi \in \Pi} \chi(\theta_0; \pi, \mathbf{z}) \sum_{B \in \pi} \left( \frac{\mu_S(\theta_0; B, \mathbf{z}) - \mu(\theta_0; B, \mathbf{z})}{\mu(\theta_0; B, \mathbf{z})} \right) \\
&\quad + \frac{1}{\delta(\theta_0; \mathbf{z})} \sum_{\pi \in \Pi} \left( \sum_{B' \in \pi} \nabla_\theta \log \mu(\theta_0; B', \mathbf{z}) \right) \chi(\theta_0; \pi, \mathbf{z}) \sum_{B \in \pi} \left( \frac{\mu_S(\theta_0; B, \mathbf{z}) - \mu(\theta_0; B, \mathbf{z})}{\mu(\theta_0; B, \mathbf{z})} \right) \\
&\quad + \frac{1}{\delta(\theta_0; \mathbf{z})} \sum_{\pi \in \Pi} \chi(\theta_0; \pi, \mathbf{z}) \sum_{B \in \pi} (\nabla_\theta \log \mu_S(\theta_0; B, \mathbf{z}) - \nabla_\theta \log \mu(\theta_0; B, \mathbf{z})) + O_{\Pr}(S^{-1}).
\end{aligned}$$

Now note that

$$\mu_S(\theta_0; B, \mathbf{z}) - \mu(\theta_0; B, \mathbf{z}) = \int a(v; \theta_0, B, \mathbf{z}) \left( (dF_{V,S}(v) - dF_V(v)) \right)$$

and, by using again the law of large numbers,

$$\begin{aligned}
&\nabla_\theta \log \mu_S(\theta_0; B, \mathbf{z}) - \nabla_\theta \log \mu(\theta_0; B, \mathbf{z}) \\
&= \frac{1}{\mu(\theta_0; B, \mathbf{z})} \int \nabla_\theta a(v; \theta_0, B, \mathbf{z}) \left( (dF_{V,S}(v) - dF_V(v)) \right) \\
&\quad - \frac{\nabla_\theta \log \mu(\theta_0; B, \mathbf{z})}{\mu(\theta_0; B, \mathbf{z})} \int a(v; \theta_0, B, \mathbf{z}) \left( (dF_{V,S}(v) - dF_V(v)) \right) + O_{\Pr}(S^{-1}).
\end{aligned}$$

We therefore deduce that

$$\nabla_\theta \ell_{1S}(\theta_0; \mathbf{z}) = \nabla_\theta \ell_1(\theta_0; \mathbf{z}) + \int \psi(v; \theta_0, \pi, \mathbf{z}) \left( (dF_{V,S}(v) - dF_V(v)) \right) + O_{\Pr}(S^{-1}).$$

Observe that, by the central limit theorem,

$$\frac{1}{n^{1/2}} \sum_{i=1}^n \nabla_\theta \ell_1(\theta_0; \mathbf{z}_i) \xrightarrow{d} \mathcal{N}(0, I_1(\theta_0))$$

and, by the Lindeberg theorem (assuming without loss of generality that $n = n_S$ is a function of $S$),

$$\frac{S^{1/2}}{n} \sum_{i=1}^{n} \int \psi(v; \theta_0, \pi, \mathbf{z}_i)\left((dF_{V,S}(v) - dF_V(v))\right)$$

$$= \frac{1}{S^{1/2}} \sum_{s=1}^{S} \left[\frac{1}{n} \sum_{i=1}^{n} \left(\psi((v_s; \theta_0, \pi, \mathbf{z}_i) - \mathbb{E}\left[\psi(V; \theta_0, \pi, \mathbf{z}_l)\right]\right)\right]$$

$$\xrightarrow{d} \mathcal{N}(0, \mathbb{V}\left(\mathbb{E}\left[\psi(V; \theta_0, \pi, \mathbf{Z})\,|V\right]\right)).$$

By using the same arguments as above, we deduce that

$$\frac{1}{n} \sum_{i=1}^{n} \nabla_{\theta}^2 \ell_{1S}(\theta_0; \mathbf{z}_i) = \frac{1}{n} \sum_{i=1}^{n} \nabla_{\theta}^2 \ell_1(\theta_0; \mathbf{z}_i) + O_{\mathrm{Pr}}\left(S^{-1/2}\right) + O_{\mathrm{Pr}}\left(n^{-1/2}S^{-1}\right)$$

$$= -I_1(\theta_0) + O_{\mathrm{Pr}}(n^{-1/2}) + O_{\mathrm{Pr}}\left(S^{-1/2}\right) + O_{\mathrm{Pr}}\left(n^{-1/2}S^{-1}\right).$$

It turns that

$$0 = \frac{1}{n^{1/2}} \sum_{i=1}^{n} \nabla_{\theta} \ell_1(\theta_0; \mathbf{z}_i) - I_1(\theta_0)\, n^{1/2}(\hat{\theta}_{n,S}^{(1)} - \theta_0)$$

$$+ \left(\frac{n}{S}\right)^{1/2} \left(\frac{S^{1/2}}{n} \sum_{i=1}^{n} \int \psi(v; \theta_0, \pi, \mathbf{z}_l)\left((dF_{V,S}(v) - dF_V(v))\right)\right)$$

$$+ O_{\mathrm{Pr}}\left(\hat{\theta}_{n,S}^{(1)} - \theta_0\right) + O_{\mathrm{Pr}}\left(\left(\frac{n}{S}\right)^{1/2}(\hat{\theta}_{n,S}^{(1)} - \theta_0)\right) + O_{\mathrm{Pr}}\left(S^{-1}(\hat{\theta}_{n,S}^{(1)} - \theta_0)\right).$$

Therefore
- if $n/S$ tends to zero, then

$$\sqrt{n}(\hat{\theta}_{n,S}^{(1)} - \theta_0) \xrightarrow{d} \mathcal{N}\left(0, I_1^{-1}(\theta_0)\right);$$

- if $n/S$ tends to infinity, then

$$\sqrt{S}(\hat{\theta}_{n,S}^{(1)} - \theta_0) \xrightarrow{d} \mathcal{N}\left(0, I_1^{-1}(\theta_0)\,\Sigma(\theta_0)\,I_1^{-1}(\theta_0)\right);$$

- if $n/S$ tends to $\varphi$, then

$$\sqrt{n}(\hat{\theta}_{n,S}^{(1)} - \theta_0) \xrightarrow{d} \mathcal{N}\left(0, I_1^{-1}(\theta_0)\left(I_1(\theta_0) + \varphi\Sigma(\theta_0)\right)I_1^{-1}(\theta_0)\right).$$

## 7.2 Proof of Proposition 8

Let

$$\mathbf{X}_{i,n/k} = \left(\frac{\mathbf{Y}_i}{\mathbf{U}_{\mathbf{Y}}(n/k)}\right)^{\boldsymbol{\alpha}} \vee \mathbf{e}$$

where $\mathbf{U}_{\mathbf{Y}}(n/k) = (U_{Y_1}(n/k), \dots, U_{Y_m}(n/k))'$ with $U_{Y_j}(n/k) = F_{Y_j}^{\leftarrow}(1 - k/n)$.

For $\hat{\mathbf{X}}_{i,n/k} \in \mathcal{A}_B$,

$$
\begin{aligned}
& \nabla_\theta \ell_{\mathbf{X}^*} \left(\hat{\theta}_k; \hat{\mathbf{X}}_{i,n/k}\right) - \nabla_\theta \ell_{\mathbf{X}^*} \left(\hat{\theta}_k; \mathbf{X}_{i,n/k}\right) \\
= \; & -\sum_{j \in B} \alpha_j \nabla^2_{\theta x_j} \ell_{\mathbf{X}^*} \left(\hat{\theta}_k; \mathbf{X}_{i,n/k}\right) X_{i,j,n/k} \left[ \left( \frac{\hat{U}_{Y,j}\left(n/k\right) - U_{Y,j}\left(n/k\right)}{U_{Y,j}\left(n/k\right)} \right) + \left( \frac{1}{\hat{\alpha}_{j,n}} - \frac{1}{\alpha_j} \right) \log X_{i,j,n/k} \right] \\
& + o_{\Pr} \left( k^{-1/2} \right).
\end{aligned}
$$

We used a first order expansion of $\nabla_\theta \ell_{\mathbf{X}^*}$ around $\boldsymbol{\alpha}$ and $\mathbf{U_Y}\left(n/k\right)$. Let, for $x_j > 1$,

$$
N^{(j)}_k \left(x_j\right) = \sum_{B \in \mathcal{P}, B \cap \{j\} \neq \varnothing} N_{B,k} \left((1,\infty) \times (1,\infty) \times \ldots \times (x_j, \infty) \times \ldots \times (1,\infty)\right) = \sum_{i=1}^{n} \mathbb{I}\{X_{i,j,n/k} > x_j\}
$$

and define

$$
W_{j,n}\left(x_j\right) = \sqrt{k} \left( \frac{1}{k} N^{(j)}_k\left(x_j\right) - x_j^{-1} \right)
$$

As for the proof of Proposition 5, it may be shown that $W_{j,n}\left(\cdot\right) \Longrightarrow W_j\left(\cdot\right)$, and by using the same arguments as in [12] we get

$$
\begin{aligned}
\sqrt{k} \left( \frac{1}{\hat{\alpha}_{j,n}} - \frac{1}{\alpha_j} \right) & \rightarrow \int_1^\infty W_j \left(x_j^{\alpha_j}\right) \frac{dx_j}{x_j} - \frac{1}{\alpha_j} W_j\left(1\right) \\
\sqrt{k} \left( \frac{\hat{U}_{Y,j}\left(n/k\right) - U_{Y,j}\left(n/k\right)}{U_{Y,j}\left(n/k\right)} \right) & \rightarrow \frac{1}{\alpha_j} W_j\left(1\right).
\end{aligned}
$$

The maximum likelihood estimator (MLE) satisfies the condition

$$
\begin{aligned}
0 & = \frac{1}{k^{1/2}} \sum_{i \in \mathcal{N}_k} \nabla_\theta \ell_{\mathbf{X}^*} \left(\hat{\theta}_k; \hat{\mathbf{X}}_{i,n/k}\right) \\
& = \frac{1}{k^{1/2}} \sum_{i \in \mathcal{N}_k} \nabla_\theta \ell_{\mathbf{X}^*} \left(\hat{\theta}_k; \mathbf{X}_{i,n/k}\right) \\
& \quad + \frac{1}{k^{1/2}} \sum_{B \in \mathcal{P}} \sum_{\hat{\mathbf{X}}_{i,n/k} \in \mathcal{A}_B} \left( \nabla_\theta \ell_{\mathbf{X}^*} \left(\hat{\theta}_k; \hat{\mathbf{X}}_{i,n/k}\right) - \nabla_\theta \ell_{\mathbf{X}^*} \left(\hat{\theta}_k; \mathbf{X}_{i,n/k}\right) \right).
\end{aligned}
$$

An expansion of the first term around $\theta_0$ gives

$$
\begin{aligned}
& \frac{1}{k^{1/2}} \sum_{i \in \mathcal{N}_k} \nabla_\theta \ell_{\mathbf{X}^*} \left(\hat{\theta}_k; \mathbf{X}_{i,n/k}\right) \\
= \; & \frac{1}{k^{1/2}} \sum_{i \in \mathcal{N}_k} \nabla_\theta \ell_{\mathbf{X}^*} \left(\theta_0; \mathbf{X}_{i,n/k}\right) + \frac{1}{k} \sum_{i \in \mathcal{N}_k} \nabla^2_\theta \ell_{\mathbf{X}^*} \left(\theta_0; \mathbf{X}_{i,n/k}\right) k^{1/2} \left(\hat{\theta}_k - \theta_0\right) + O_{\Pr} \left( k^{1/2} (\hat{\theta}_k - \theta_0)^2 \right)
\end{aligned}
$$

Note that i)

$$
\frac{1}{k} \sum_{i \in \mathcal{N}_k} \nabla^2_\theta \ell_{\mathbf{X}^*} \left(\theta_0; \mathbf{X}_{i,n/k}\right) \xrightarrow{\Pr} V_*(\theta_0; \mathbf{e}) I_{\mathbf{X}^*}\left(\theta_0\right)
$$

ii)

$$
\frac{1}{k^{1/2}} \sum_{i \in \mathcal{N}_k} \nabla_\theta \ell_{\mathbf{X}^*} \left(\theta_0; \mathbf{X}_{i,n/k}\right) \xrightarrow{d} \sum_{B \in \mathcal{P}} \int_{\mathcal{A}_B} \nabla_\theta \ell_{\mathbf{X}^*} \left(\theta_0; \mathbf{x}\right) W_B\left(d\mathbf{x}_B\right).
$$

Now

$$\frac{1}{k^{1/2}} \sum_{B \in \mathcal{P}} \sum_{\hat{\mathbf{X}}_{i,n/k} \in \mathcal{A}_B} \left( \nabla_\theta \ell_{\mathbf{X}^*} \left( \hat{\theta}_k; \hat{\mathbf{X}}_{i,n/k} \right) - \nabla_\theta \ell_{\mathbf{X}^*} \left( \hat{\theta}_k; \mathbf{X}_{i,n/k} \right) \right)$$

$$= - \sum_{B \in \mathcal{P}} \sum_{j \in B} \alpha_j \left( \frac{\hat{U}_{Y,j}\left(n/k\right) - U_{Y,j}\left(n/k\right)}{U_{Y,j}\left(n/k\right)} \right) \left( \sum_{\mathbf{X}_{i,n/k} \in \mathcal{A}_B} \nabla^2_{\theta x_j} \ell_{\mathbf{X}^*} \left( \hat{\theta}_k; \mathbf{X}_{i,n/k} \right) \mathbf{X}_{i,j,n/k} \right)$$

$$\quad - \sum_{B \in \mathcal{P}} \sum_{j \in B} \alpha_j \left( \frac{1}{\hat{\alpha}_{j,k,n}} - \frac{1}{\alpha_j} \right) \left( \sum_{\mathbf{X}_{i,n/k} \in \mathcal{A}_B} \nabla^2_{\theta x_j} \ell_{\mathbf{X}^*} \left( \hat{\theta}_k; \mathbf{X}_{i,n/k} \right) \mathbf{X}_{i,j,n/k} \log \mathbf{X}_{i,j,n/k} \right) + o_{\mathrm{Pr}}\left(1\right).$$

Note also that

$$\sum_{B \in \mathcal{P}} \sum_{j \in B} \alpha_j \left( \frac{\hat{U}_{Y,j}\left(n/k\right) - U_{Y,j}\left(n/k\right)}{U_{Y,j}\left(n/k\right)} \right) \left( \sum_{\mathbf{X}_{i,n/k} \in \mathcal{A}_B} \nabla^2_{\theta x_j} \ell_{\mathbf{X}^*} \left( \hat{\theta}_k; \mathbf{X}_{i,n/k} \right) \mathbf{X}_{i,j,n/k} \right)$$

$$= \sum_{j=1}^{m} \alpha_j \left( \frac{\hat{U}_{Y_j}\left(n/k\right) - U_{Y,j}\left(n/k\right)}{U_{Y,j}\left(n/k\right)} \right) \sum_{B \in \mathcal{P}, B \cap \{j\} \neq \varnothing} \sum_{\mathbf{X}_{i,n/k} \in \mathcal{A}_B} \nabla^2_{\theta x_j} \ell_{\mathbf{X}^*} \left( \hat{\theta}_k; \mathbf{X}_{i,n/k} \right) \mathbf{X}_{i,j,n/k}$$

and

$$\sum_{B \in \mathcal{P}} \sum_{j \in B} \alpha_j \left( \frac{1}{\hat{\alpha}_{j,k,n}} - \frac{1}{\alpha_j} \right) \left( \sum_{\mathbf{X}_{i,n/k} \in \mathcal{A}_B} \nabla^2_{\theta x_j} \ell_{\mathbf{X}^*} \left( \hat{\theta}_k; \mathbf{X}_{i,n/k} \right) \mathbf{X}_{i,j,n/k} \log \mathbf{X}_{i,j,n/k} \right)$$

$$= \sum_{j=1}^{m} \alpha_j \left( \frac{1}{\hat{\alpha}_{j,k,n}} - \frac{1}{\alpha_j} \right) \sum_{B \in \mathcal{P}, B \cap \{j\} \neq \varnothing} \sum_{\mathbf{X}_{i,n/k} \in \mathcal{A}_B} \nabla^2_{\theta x_j} \ell_{\mathbf{X}^*} \left( \hat{\theta}_k; \mathbf{X}_{i,n/k} \right) \mathbf{X}_{i,j,n/k} \log \mathbf{X}_{i,j,n/k}.$$

Moreover

$$\frac{1}{k} \sum_{B \in \mathcal{P}, B \cap \{j\} \neq \varnothing} \sum_{\mathbf{X}_{i,n/k} \in \mathcal{A}_B} \nabla^2_{\theta x_j} \ell_{\mathbf{X}^*} \left( \hat{\theta}_k; \mathbf{X}_{i,n/k} \right) \mathbf{X}_{i,j,n/k}$$

$$= \frac{N_k^{(j)}\left(1\right)}{k} \frac{1}{N_k^{(j)}\left(1\right)} \sum_{B \in \mathcal{P}, B \cap \{j\} \neq \varnothing} \sum_{\mathbf{X}_{i,n/k} \in \mathcal{A}_B} \nabla^2_{\theta x_j} \ell_{\mathbf{X}^*} \left( \hat{\theta}_k; \mathbf{X}_{i,n/k} \right) \mathbf{X}_{i,j,n/k}$$

$$\xrightarrow{\mathrm{Pr}} \mathbb{E}\left[ \nabla^2_{\theta x_j} \ell_{\mathbf{X}^*} \left( \theta_0; \mathbf{X}_{i,n/k} \right) X_j^* \Big| X_j^* > 1 \right] = \omega_j$$

and

$$\frac{1}{k} \sum_{B \in \mathcal{P}, B \cap \{j\} \neq \varnothing} \sum_{\mathbf{X}_{i,n/k} \in \mathcal{A}_B} \nabla^2_{\theta x_j} \ell_{\mathbf{X}^*} \left( \hat{\theta}_k; \mathbf{X}_{i,n/k} \right) \mathbf{X}_{i,j,n/k} \log \mathbf{X}_{i,j,n/k}$$

$$\xrightarrow{\mathrm{Pr}} \mathbb{E}\left[ \nabla^2_{\theta x_j} \ell_{\mathbf{X}^*} \left( \theta_0; \mathbf{X}_{i,n/k} \right) X_j^* \log X_j^* \Big| X_j^* > 1 \right] = \psi_j \ .$$

It follows that

$$\sqrt{k} \sum_{j=1}^{m} \alpha_j \left( \frac{\hat{U}_{Y_j}\left(n/k\right) - U_{Y,j}\left(n/k\right)}{U_{Y,j}\left(n/k\right)} \right) \frac{1}{k} \sum_{B \in \mathcal{P}, B \cap \{j\} \neq \varnothing} \sum_{\mathbf{X}_{i,n/k} \in \mathcal{A}_B} \nabla^2_{\theta x_j} \ell_{\mathbf{X}^*} \left( \hat{\theta}_k; \mathbf{X}_{i,n/k} \right) \mathbf{X}_{i,j,n/k}$$

$$\to \sum_{j=1}^{m} \omega_j W_j\left(1\right)$$

and

$$\sqrt{k} \sum_{j=1}^{m} \alpha_j \left( \frac{1}{\hat{\alpha}_{j,k,n}} - \frac{1}{\alpha_j} \right) \frac{1}{k} \sum_{B \in \mathcal{P}, B \cap \{j\} \neq \varnothing} \sum_{\mathbf{X}_{i,n/k} \in \mathcal{A}_B} \nabla^2_{\theta x_j} \ell_{\mathbf{X}^*} \left( \hat{\theta}_k; \mathbf{X}_{i,n/k} \right) \mathbf{X}_{i,j,n/k} \log \mathbf{X}_{i,j,n/k}$$

$$\rightarrow \sum_{j=1}^{m} \alpha_j \psi_j \left( \int_{1}^{\infty} W_j \left( z_j^{\alpha_j} \right) \frac{dz_j}{z_j} - \frac{1}{\alpha_j} W_j \left( 1 \right) \right).$$

This concludes the proof.